\documentclass[prd,showpacs,twocolumn]{revtex4-1}
\usepackage{slashed}
\usepackage{mathrsfs}
\usepackage{amsfonts}
\usepackage{amsmath}
\usepackage{amssymb}
\usepackage{revsymb}
\usepackage{graphicx}
\usepackage{mathrsfs}
\usepackage{bm}
\usepackage{psfrag}
\usepackage{color}
\usepackage{hyperref}
\usepackage{natbib}

\usepackage[usenames,dvipsnames]{xcolor}

\newcommand{\bea}{\begin{eqnarray}} 
\newcommand{\eea}{\end{eqnarray}} 
\newcommand{\eps}{\varepsilon}

\newcommand{\trm}[1]{\textrm{#1}}

\newcommand{\figref}[1]{Fig. \ref{#1}}
\newcommand{\figrefa}[1]{Fig. \ref{#1}a}
\newcommand{\figrefb}[1]{Fig. \ref{#1}b}

\newcommand{\eqnref}[1]{Eq. (\ref{#1})}
\newcommand{\eqnrefs}[2]{Eqs. (\ref{#1}) and (\ref{#2})}
\newcommand{\eqnreft}[2]{Eqs. (\ref{#1}-\ref{#2})}
\newcommand{\sxnref}[1]{Sec. \ref{#1}}

\newcommand{\tr}{\trm{tr}\,}
\newcommand{\Ecr}{E_{\trm{cr}}}
\newcommand{\psibar}{\overline{\psi}}

\newcommand{\sk}{\slashed{\varkappa}}
\newcommand{\sA}{\slashed{A}}
\newcommand{\vkap}{\varkappa}
\newcommand{\vphi}{\varphi}

\newcommand{\defto}{=}
\newcommand{\Ai}{\trm{Ai}}
\newcommand{\ra}{\rightarrow}
\newcommand{\la}{\leftarrow}
\newcommand{\Iod}{\mathcal{I}^{(1)}_{\trm{d}}}
\newcommand{\Iox}{\mathcal{I}^{(1)}_{\times}}

\begin{document}
\title{Trident pair production in a constant crossed field}
\author{B. \surname{King}}
  \email{ben.king@physik.uni-muenchen.de}
\author{H. \surname{Ruhl}}
\affiliation{Arnold Sommerfeld Center for Theoretical Physics, \\ Ludwig-Maximilians-Universit\"at M\"unchen,
    Theresienstra\ss e 37, 80333 M\"unchen, Germany}


\date{\today}
\begin{abstract}
We isolate the two-step mechanism involving a real intermediate photon from the one-step mechanism involving a virtual photon for the trident process in a constant crossed field. The two-step process is shown to agree with an integration over polarised sub-processes. At low to moderate quantum non-linearity parameter, the one-step process is found to be suppressed. When the parameter is large, the two decay channels are comparable if the field dimensions are not much greater than the formation length. 
\end{abstract}
\maketitle

\section{Introduction}
Partly due to experiments that have measured them, and partly due to theoretical proposals to observe them, higher-order quantum-electrodynamical processes in external fields have recently gained much attention in the literature. Theoretical results for two-photon non-linear Compton scattering in a pulsed laser field \cite{mackenroth13, seipt12, *loetstedt09a, *loetstedt09b} have shown in particle spectra a much richer physics of higher-order processes compared to tree-level versions. Recent attention has also been focused on the trident process in an external field, which is essentially lowest-order fermion-seeded pair creation, $e^{\pm}\to e^{\pm} + e^{+}e^{-}$, where $e^{+}$ represents a positron and $e^{-}$ an electron. Part of the trident process was measured in the landmark  E-144 experiment at  SLAC \cite{burke97, bamber99}, which still more than a decade later is being analysed by theorists \cite{hu10,ilderton11} despite higher-order processes, including trident, having first been studied some years 
before \cite{baier72, ritus72, narozhny77} (a review of strong-field effects in quantum electrodynamics (QED) can be found in \cite{ritus85, dipiazza12,*ehlotzky09,*marklund_review06}).

In light of several plans to construct the next-generation of high-intensity lasers \cite{ELI_SDR,*HiPER_TDR,*XCELS_SDR,*GEKKO_SDRb}, there has been much activity in attempting to simulate relativistic plasmas that include strong-field QED effects \cite{ridgers12b,*ridgers12,*elkina11,*ruhl10}. Due to their complexity and the current lack of a consistent framework for including classical and quantum effects alongside one another, approximations must be employed (although some non-perturbative strong-field QED simulation methods for systems with fewer particles are also being currently developed \cite{hebenstreit13, ilderton13c}). The current letter is motivated on the one hand by the need to justify approximating higher-order QED processes by chains of tree-level processes in  simulation-based approaches, and on the other by an enquiry into the physics of the trident process in an external field. This study complements the numerical approach of \cite{hu10} that analysed the weakly nonlinear regime in E-144 using a monochromatic plane wave background modified to take into account finite interaction time, the lucid general theoretical outline of \cite{ilderton11} and the 
expression for the total rate in a constant crossed field neglecting exchange terms derived in \cite{baier72, ritus72}. By deriving an analytical expression for the trident process in a constant crossed field, we will separate off in an unambiguous way, the two-step process, measured in E-144 in a laser pulse, of a real photon produced via non-linear Compton scattering decaying into an electron-positron pair ($e^{\pm} \to e^{\pm} + \gamma$, $\gamma \to e^{+}e^{-}$, where $\gamma$ represents a photon). Moreover, we will show that the two-step process is exactly given by a sum over intermediate photon polarisation of each tree-level sub-process integrated over the photon lightfront momentum and that for small quantum non-linearity parameter, the one-step process involving a virtual photon suppresses the total rate. Furthermore, we compare the relative importance of the two decay channels and comment on the measurability.

The paper is organised as follows. We begin by highlighting important points in the derivation of the trident process in a constant crossed field, relegating technical albeit standard steps to the appendix. The two-step contribution is analysed and compared to combining tree-level rates and then the remaining, nominatively ``one-step'' contribution is analysed and compared to the Weizs\"acker-Williams approximation. The total creation probability is then studied, the measurability of each process commented on, the results discussed and the paper concluded.

\section{Probability derivation outline}
\begin{figure}[!ht]
\centering\noindent
  \includegraphics[draft=false, width=4.0cm]{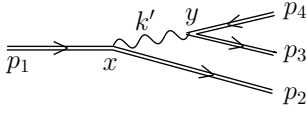}
\caption{The Feynman diagram for one term of the trident process in a plane wave (the other is given by the substitution $p_{2}\leftrightarrow p_{3}$).}
\label{fig:feyndiag}
\end{figure}

 A diagram of the considered trident process is given in \figref{fig:feyndiag}, where double lines indicate fermions dressed in the external field, which has a vector potential $A^{\mu}(\varphi)$, phase $\varphi=\vkap x$ and wavevector $\vkap$, satisfying $\vkap A = \vkap^{2} = 0$.  Following standard Feynman rules (see e.g. \cite{landau4}), in a system of units $c=\hbar=1$ with the fine-structure constant $\alpha = e^{2}$, for positron charge and mass $e>0$, $m$, the scattering matrix for this trident process is given by:
\bea
S_{fi}\!\!&=&\!\!\alpha\!\!\int\! d^{4}x \,d^4y \,\, \psibar_{2}(x)\gamma^{\mu}\psi_{1}(x)G_{\mu\nu}(x-y)\psibar_{3}(y)\gamma^{\nu}\psi^{+}_{4}(y) \nonumber \\ 
&& \qquad\qquad - (p_{2}\leftrightarrow p_{3}), \label{eqn:Sfi1}
\eea
where the electron in, electron out and positron out wavefunctions in the field of a plane wave $\psi$, $\psibar$, $\psi^{+}$ are given by Volkov states \cite{volkov35},  $G_{\mu\nu}(x-y)$ is the photon propagator and $(p_{2}\leftrightarrow p_{3})$ refers to an exchange of $p_{2}$ and $p_{3}$ in the first term of $S_{fi}$ and $S_{fi} =\overrightarrow{S_{fi}} - \overleftarrow{S_{fi}}$. The second term must be subtracted due to exchange symmetry as the two outgoing electrons are indistinguishable (Pauli's principle). To avoid ambiguity arising from the exchange term, differential rates will be in $p^{\prime}$, $p_{-}$ and $p_{+}$, which refer to the scattered electron, created electron and positron momentum respectively.

Let us fix the co-ordinate system by defining $\vkap^{\mu} = \vkap^{0}(1,0,0,1)^{\mu}$, $A^{\mu} = A(\varphi)(0,1,0,0)^{\mu}$. Then focusing on just $\overrightarrow{S_{fi}}$ (the calculation for $\overleftarrow{S_{fi}}$ is analogous), using the definition of the objects in \eqnref{eqn:Sfi1} and Fourier-transforming both vertices $x$ and $y$, one arrives at \footnote{More detailed derivation steps and definitions are given in the appendix}:
\bea
\overrightarrow{S_{fi}} &=& (2\pi)^{2}\!\alpha\!\!\int\!\! drds\,\delta^{(4)}(\Pi)\Gamma^{\mu}(r)\,\frac{1}{k'^{2}+i\varepsilon}\Big|_{k'=k'_{\ast}}\!\!\!\!\Delta_{\mu}(s), \label{eqn:Sfi3}
\eea
where $\Pi= p_{2}+p_{3}+p_{4}-p_{1} -(r+s)\vkap $, $k'$ is the photon wavevector, $k'_{\ast}=p_{1}-p_{2} +r\vkap$ and $\Gamma^{\mu}(r)$ and $\Delta_{\mu}(s)$ are functions of variables at the first and second vertices respectively. It has been shown that the Fourier-transformation variables $r$ and $s$ are equivalent to the number of external-field photons, when the background is an infinite plane wave \cite{ritus85}. \\

A constant crossed field background $A^{\mu}(\vphi) = a^{\mu} \vphi$ is interesting, first because many integrals can be performed analytically facilitating physical interpretation, second, that integration is computationally sufficiently cheap that rates could feasibly be added to simulations and third that predictions in a constant crossed field are often a good approximation to in an arbitrary background field. When one considers that a general strong-field QED process can depend on four gauge- and relativistic- invariants \cite{ilderton09}
\begin{align}
\begin{split}
\xi = \frac{e^{2} p_{\mu}T^{\mu\nu}p_{\nu}}{m^{2} (\vkap p)^{2}};\quad \chi = \frac{e \sqrt{|p_{\mu}F^{\mu\nu}|^{2}}}{m^{3}};\\
\quad \mathcal{F}=\frac{e^{2} F_{\mu\nu}F^{\mu\nu}}{4 m^{4}}; \quad \mathcal{G}=\frac{e^{2} F^{\ast}_{\mu\nu}F^{\mu\nu}}{4 m^{4}},
\end{split}
\end{align}
where $T^{\mu\nu}$ and $F^{\mu\nu}$ are the energy-momentum and Faraday tensors and $\xi$ and $\chi$ the classical and quantum non-linearity parameters, it is a common argument \cite{ritus85} that if $\xi \gg 1$ (equivalent to process formation lengths being much smaller than the external field wavelength) the external field can be considered constant during the process, and if $\mathcal{F}, \mathcal{G} \ll \chi^{2}, 1$, then probabilities $P$ are well-approximated by those in a constant crossed field $P(\chi, \mathcal{F}, \mathcal{G})\approx P(\chi, 0,0)$. The classically non-linear regime $\xi\gg1$ is fulfilled by the most intense lasers \cite{yanovsky08}, as are $\mathcal{F},\mathcal{G}\ll1$.
\newline

The probability of the trident process can be calculated by performing the trace average over spin states (achieved using the package {\tt{Feyncalc}} \cite{feyncalc}) and integrating over the outgoing degrees of freedom (a factor $1/2$ removes double-counting from identical final particles), $P = (1/4)\prod_{j=2}^{4}[V\int d^{3}p_{j}/(2\pi)^{3}] \tr|S_{fi}|^{2}$, where $V$ is the system volume. When the momentum-conserving delta-function in \eqnref{eqn:Sfi3} is squared, a factor in the denominator of a \emph{formation phase length}, $\Delta\varphi_{+}$, is generated:
\bea
\delta(r+s-(r'+s'))\Big|_{r+s=r'+s'}= \Delta\varphi_{+}/2\pi,
\eea
where $\Delta\varphi_{+} = \int d\varphi_{+}$, $\varphi_{\pm}=\varphi_{x}\pm\varphi_{y}$ is assumed finite and $\varphi_{z}=z\varkappa$, $z\in\{x,y\}$ (more detail of this step is given in \eqnreft{eqn:form1eqS}{eqn:form1eqE}). The formation phase length can be related to particle momenta by calculating the position of the (real) saddle-point in the phase of $\overrightarrow{S_{fi}}$, $\varphi^{\ast}_{+}=\varphi^{\ast}_{x}+\varphi^{\ast}_{y}$ and then associating $\Delta\varphi^{\ast}_{+} = \int d\varphi^{\ast}_{+}$ analogous to tree-level calculations \cite{ritus85}, explained in more detail in \sxnref{app:temp}. Replacing phase lengths $\Delta\varphi$ with coherence intervals $\Delta\varphi^{\ast}$ is a key part of this approach and they will be used interchangeably.

As the rate is proportional to $|S_{fi}|^{2}=|\overrightarrow{S_{fi}}+\overleftarrow{S_{fi}}|^{2}$, we note interference between exchange terms arises. In \sxnref{app:crossremoval}, it is argued that this interference is negligible when the field dimensions are much larger than the formation length. This is the only part neglected as we proceed with $P \approx (\overrightarrow{P} +\overleftarrow{P})/2$. Moreover, $p_{2}\leftrightarrow p_{3}$ is a symmetry of the remaining integrand, permitting us to define $P=\overrightarrow{P}=\overleftarrow{P}$.

Using the definition of $P$ given above with the replacement $S_{fi}\to\overrightarrow{S_{fi}}$, one then has:
\begin{gather}
\overrightarrow{P} = \frac{\alpha^{2}\xi^{2}m^{2}}{ 2^{7}(\vkap^{0})^{3} (\vkap p_{1})}\int\!\frac{dp^{-}_{2}dp^{-}_{3}dp^{2}_{2}dp^{2}_{3}}{4\pi p_{1}^{-}p_{2}^{-}p_{3}^{-}p_{4}^{-}} \frac{\theta(p_{2}^{-})\theta(p_{3}^{-})}{p_{1}^{-}-p_{2}^{-}}\mathcal{J}\\ 
\mathcal{J} = \int \frac{d\varphi^{\ast}_{+}\, d\varphi^{\ast}_{-}}{2\pi^{2}} \tr \left|\int\!dr\,\frac{\Gamma^{\mu}(r+r_{\ast})\Delta_{\mu}(s_{\ast} -r)}{r+i\eps}\right|^{2}_{\textrm{nn}} 
\end{gather}
where the integral in $d^{3}p_{4}$ has already been performed, $\xi=m\chi_{E}/\varkappa^{0}$, the instruction $\textrm{nn}$ means that all normalisations of the form $(2Vp_{j}^{0})^{-1/2}$ for $j \in \{1,2,3,4\}$ have been removed, $r_{\ast}= (p_{1}-p_{2})^{2}/2\vkap(p_{2}-p_{1})$ and $s_{\ast}=[(p_{2}+p_{3}+p_{4})^{2}-m^{2}]/2p_{1}\vkap -r_{\ast}$ (the steps to arrive at this formula are detailed in the appendix). Focusing on the $r$-integration, we can write:
\bea
\mathcal{J} &=& \frac{1}{2\pi^{2}}\int d\varphi^{\ast}_{+}\, d\varphi^{\ast}_{-}\,\left|\int\!dr~\frac{\mbox{e}^{i\varphi_{-}^{\ast}r}F(r)}{r+i\eps}\right|^{2},
\eea
where $F(r) \in \mathbb{C}^{\infty}$. A crucial step is how to deal with the integration over the photon propagator. As commented in \cite{ilderton11}, using the Sokhotsky-Plemelj formula \cite{heitler60}:
\bea
\int_{-\infty}^{\infty} \!dr~ \frac{F(r)}{r\pm i \eps} = \mp i \pi F(0) + \hat{\mathcal{P}}\int_{-\infty}^{\infty}\!\!dr ~\frac{F(r)}{r}, \label{eqn:Plemelj}
\eea
where $\hat{\mathcal{P}}$ refers to taking the Cauchy principal value of the integral, $P$ might thought to be split into \emph{real} and \emph{virtual} parts, for which the photon is on-shell ($k^{2}=0$) and off-shell, corresponding to the first and second terms in \eqnref{eqn:Plemelj} respectively. However, for a constant crossed field at least, these two terms correspond only to a $\delta$ \emph{function} and a \emph{principal value} part. Using \eqnref{eqn:Plemelj}, performing the principal values first, $\mathcal{J}$ can be shown to be equal to
\begin{gather}
\mathcal{J} = \mathcal{J}^{(2)} + \mathcal{J}^{(1)}_{\times} + \mathcal{J}^{(1)}_{\trm{d}}\label{eqn:Jint}\\
\mathcal{J}^{(2)} = 2|F(0)|^{2}\int d\varphi^{\ast}_{+}\, d\varphi^{\ast}_{-}\,\theta(-\varphi_{-}^{\ast})\\
\widetilde{\mathcal{J}}^{(1)}_{\times} = \frac{F(0)}{\pi}\int\!d\varphi^{\ast}_{+}\int_{0}^{\infty} \!dr~\frac{F^{\ast}(r)+F^{\ast}(-r)-2F^{\ast}(0)}{r^{2}}\\
\mathcal{J}^{(1)}_{\trm{d}} = \frac{1}{\pi}\int\!d\varphi^{\ast}_{+}\int\!dr~\frac{|F(r)-F(0)|^{2}}{r^{2}},
\end{gather}
where $\mathcal{J}^{(1)}_{\times} = 2\,\trm{Re}\,\widetilde{\mathcal{J}}^{(1)}_{\times}$, $\trm{Re}$ is the real part and $\theta(\cdot)$ is the Heaviside theta function. $\mathcal{J}^{(1)}_{\trm{d}}$ and $\mathcal{J}^{(1)}_{\times}$ denote the direct one-step and one-step-two-step interference integrals. Since $F$ most often has a maximum at $r=0$ for the most important dynamical regions, it can also be noted that the interference term is effectively negative. We return to this point in the discussion of the one-step process. Recognising that if $\varphi^{\ast}_{x},\varphi^{\ast}_{y}\in[\varphi_{0},\varphi]$ then $2^{-1}\int d\varphi^{\ast}_{+} \int d\varphi^{\ast}_{-}\theta(-\varphi^{\ast}_{-})$ is equivalent to $\int^{\varphi}_{\varphi_{0}} d\varphi^{\ast}_{x}\int_{\varphi_{0}}^{\varphi^{\ast}_{x}} d\varphi^{\ast}_{y}$ and $\mathcal{J}^{(2)}$ forms a two-step process. Although there are ostensibly two phase regions $\varphi_{x}^{\ast}$ and $\varphi_{y}^{\ast}$, in a constant crossed field, the two-step 
phase factor is trivially related to the total phase region through
\bea
\int^{\varphi}_{\varphi_{0}} d\varphi^{\ast}_{x}\int_{\varphi_{0}}^{\varphi^{\ast}_{x}} d\varphi^{\ast}_{y} =  \frac{(\Delta \varphi)^{2}}{2},
\eea
where $\Delta\varphi=\varphi-\varphi_{0}$ is the total phase difference between the beginning and end of the process in the external field, thereby measuring its extent. What remains in $P$ are terms proportional to $2^{-1}\int d\varphi_{+}^{\ast}$, equal to $\int_{\varphi_{0}}^{\varphi}\!d\varphi'=\Delta\varphi$, which we deem accordingly a one-step process. The Heaviside theta function in the two-step process preserves causality, ensuring that pair-creation from a photon occurs after photon emission from non-linear Compton scattering. An important point is that this theta function is generated from terms in both the $\delta$ function and principal value part of the photon propagator and so the principal value part also plays a key role in the calculation of the two-step process involving a real photon. Even in the remaining one-step terms in \eqnref{eqn:Jint} that comprise a cross-term $\mathcal{I}^{(1)}_{\times}$ and a direct-term $\mathcal{I}^{(1)}_{\trm{d}}$, it transpires that
\bea
\mathcal{I}^{(1)}_{\times}(r=0) + \mathcal{I}^{(1)}_{\trm{d}}(r=0)\neq 0.
\eea 
So as the two-step term contains a contribution from the principal value part of the propagator so does the one-step term contain a non-zero on-shell contribution. Overall causality can be seen to be observed without having to split up the propagator into $\delta$ function and principal value parts, by calculating the $r$-integral early on in $S_{fi}$ using \cite{ildertonpc}
\bea
\int_{-\infty}^{\infty}\!dr\,\frac{1}{r+i\eps}\,\mbox{e}^{ir(\varphi_{x}-\varphi_{y})} =-2\pi i\, \theta(\varphi_{y}-\varphi_{x}).\label{eqn:AIPC}
\eea\\
We can then write the total probability in terms of the two-step and one-step probabilities $P=P^{(2)}+P^{(1)}$, where $P^{(1)} = P^{(1)}_{\trm{d}} + P^{(1)}_{\times}$ comprises ``pure'' one-step and one-step-two-step cross terms.

\section{Two-step process}
From the six original outgoing momentum integrals, due to the symmetry in the $1$- (electric-field) direction, four integrals remain. As we neglect mixing between direct and exchange terms, each integral in the $2$-direction can be factorised into a Compton-scattering vertex part and a pair-creation vertex part. Some useful Airy integrals that were derived from existing results in the literature are given in \sxnref{app:Airy}, which facilitated the $p_{2}^{2}$ and $p_{3}^{2}$ integrations, giving for $P^{(2)}$ a final double-integral which can be written as a product of a spacetime-dependent phase length squared and a dynamical part dependent on relativistic and gauge-invariant quantities $\chi_{j}=\chi_{E}(p_{j}^{0}-p_{j}^{3})$, $\chi_{E}=E/\Ecr$ and $\Ecr=m^{2}/e$,
\bea
P^{(2)} &=& (\xi\Delta\varphi)^{2}\,\mathcal{I}^{(2)}/2\nonumber \\	
\mathcal{I}^{(2)} & = & \frac{\alpha^{2}}{\chi_{1}^{2}}\int \frac{d\chi_{2}d\chi_{3}\,\theta(\chi_{1}-\chi_{2}-\chi_{3})\,\mathcal{A}^{(2)}}{(\chi_{1}-\chi_{2})^{2}}, \label{eqn:P22} 
\eea
where $\mathcal{I}^{(2)}=\mathcal{I}^{(2)}(\chi_{1})$ and $\mathcal{A}^{(2)} =\mathcal{A}^{(2)}(\chi_{1},\chi_{2},\chi_{3})$ given by
\bea
\mathcal{A}^{(2)} &=& \Ai_{1}\left[\mu_{2}^{2/3}\right]\!\!\Ai_{1}\left[\mu_{3}^{2/3}\right]+ a_{2}\,\Ai'\left[\mu_{2}^{2/3}\right]\!\!\Ai_{1}\left[\mu_{3}^{2/3}\right] \nonumber \\
&&+a_{3}\,\Ai_{1}\left[\mu_{2}^{2/3}\right]\!\!\Ai'\left[\mu_{3}^{2/3}\right]+a_{4}\,\Ai'\left[\mu_{2}^{2/3}\right]\!\!\Ai'\left[\mu_{3}^{2/3}\nonumber\right] \nonumber\\
a_{2}&=&\mu_{2}^{1/3}(\chi_{1}^{2}+\chi_{2}^{2})/(\chi_{1}-\chi_{2})\nonumber\\
a_{3}&=&-\mu_{3}^{1/3}[(\chi_{1}-\chi_{2}-\chi_{3})^{2}+ \chi_{3}^{2}]/(\chi_{1}-\chi_{2})\nonumber \\
a_{4}&=&-(\mu_{2}\mu_{3})^{1/3}\big[\chi_{1}^4 - 2\chi_{1}^{3}(\chi_{2} + \chi_{3}) +  \chi_{1}\chi_{2}(-2\chi_{2}^2 \nonumber \\
&& - \chi_{2}\chi_{3} + \chi_{3}^2)  +\chi_{1}^{2}(2\chi_{2}^2 + \chi_{2}\chi_{3} + 2\chi_{3}^2) + \nonumber\\
&&   \chi_{2}^{2}(\chi_{2}^2 + 2\chi_{2}\chi_{3} + 2\chi_{3}^2)\big]/(\chi_{1}-\chi_{2})^{2}  
\eea
\begin{gather}
\mu_{2} = \frac{\chi_{1}-\chi_{2}}{\chi_{1}\chi_{2}}, \qquad
\mu_{3} = \frac{\chi_{1}-\chi_{2}}{(\chi_{1}-\chi_{2}-\chi_{3})\chi_{3}},\label{eqn:mu}
\end{gather}
where $\Ai$ is the Airy function \cite{olver97}, $\Ai'$, its differential and $\Ai_{1}(x)=\int_{x}^{\infty}\Ai(y)dy$. The phase formation length factor can be written in a variety of ways:
\bea
\xi\Delta\varphi = \frac{m \chi_{E} \Delta\varphi}{\vkap^{0}}=\frac{L}{L_{\ast}},
\eea
where $L$ is the extent of the external field and $L_{\ast}= \lambdabar/\chi_{E}$ the formation length, where $\lambdabar=1/m$ is the reduced Compton wavelength. It is quite logical that the formation length should be of this form when one considers over what extent the work done by the field is sufficient to create a pair $L_{\ast}=m/eE=\lambdabar/\chi_{E}$.
\newline

Further support that the splitting off of the two step-process is unambiguous in a constant crossed field is found upon comparison with the ``product approach'' of integrating the tree-level processes of non-linear Compton scattering (quantities denoted by subscript $_{\gamma}$) and photon-seeded pair creation (subscript $_{e}$) over the intermediate photon lightfront momenta and summing over the photon polarisation, $l$. The differential rates for the sub-process are given by \cite{king13a, nikishov64, *kibble64}:
\bea
\mathcal{I}_{\gamma, l} &=& \frac{-\alpha}{\chi_{1}^{2}} \int^{\chi_{1}}_{0}\!\!d\chi_{k} \left\{\left[\frac{2\pm1}{z_{\gamma}}+\chi_{k}z_{\gamma}^{\frac{1}{2}}\right]\Ai'(z_{\gamma}) + \Ai_{1}(z_{\gamma})\right\}\nonumber\\
\mathcal{I}_{e,l} &=& \frac{\alpha}{\chi_{k}^{2}} \int^{\chi_{k}}_{0}\!\!d\chi_{3} \left\{\left[\frac{2\pm1}{z_{e}}-\chi_{k}z_{e}^{\frac{1}{2}}\right]\Ai'(z_{e}) + \Ai_{1}(z_{e})\right\},\nonumber \\ \label{eqn:substeps}
\eea
where $z_{\gamma}=(\chi_{k}/\chi_{1}(\chi_{1}-\chi_{k}))^{2/3}=\mu_{2}^{2/3}$, $z_{e}=(\chi_{k}/\chi_{3}(\chi_{k}-\chi_{3}))^{2/3}= \mu_{3}^{2/3}$, and $\pm$ refers to transverse polarisations $l=1,2$. The probability in the product approach is $P_{\gamma e} = (\xi\Delta\varphi)^{2}\mathcal{I}_{\gamma e}/2$ where 
\bea
\mathcal{I}_{\gamma e} = \frac{1}{2}\sum_{l=1}^{2}\int_{0}^{\chi_{1}}\!\!d\chi_{k}~\frac{\partial\mathcal{I}_{\gamma,l}}{\partial\chi_{k}}\,\mathcal{I}_{e,l}.
\eea
By comparison of \eqnref{eqn:substeps} and \eqnref{eqn:P22}, one can observe that $\mathcal{I}^{(2)}=\mathcal{I}_{\gamma e}$. This agrees with previous analyses performed for the total trident rate in a constant crossed field by analysing the polarisation operator correction to the Volkov propagator in a constant crossed field \cite{baier72, ritus72}.
\newline

\begin{figure}[!h]
\centering\noindent
  \includegraphics[draft=false, width=8.6cm]{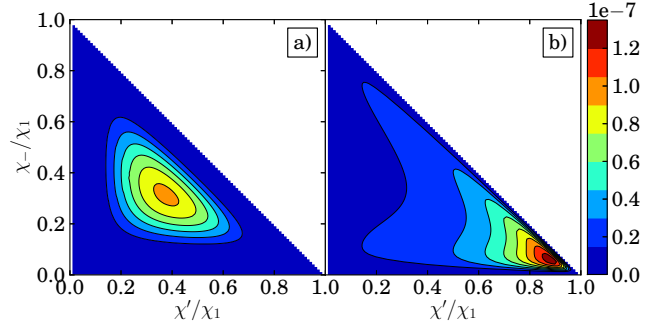}
\caption{The differential rate of the two-step process $\partial^{2}\mathcal{I}^{(2)}/\partial \chi^{\prime}\partial\chi_{-}$ for incoming fermion quantum non-linearity parameter a) $\chi_{1}=1$ and b) $\chi_{1}=10$. As $\chi_{1}$ increases above $1$, the probability becomes skewed around large $\chi^{\prime}$ and small $\chi_{-}$.} \label{fig:R22diffrate}
\end{figure}

The differential rate $\partial^{2} \mathcal{I}^{(2)}/\partial \chi^{\prime}\partial \chi_{-}$ ($\chi^{\prime}$, $\chi_{-}$ and $\chi_{+}$ refer to the scattered electron, created electron and positron $\chi$ factors respectively) is plotted in \figref{fig:R22diffrate} for incoming quantum non-linearity parameter $\chi_{1}=1$ and $\chi_{1}=10$. It was found that for $\chi_{1}\leq 1$, the differential rate is symmetric in $\chi^{\prime}$ and $\chi_{-}$, becoming more sharply peaked around the maximum at $\chi^{\prime}=\chi_{-}=\chi_{+}=\chi_{1}/3$ as the initial $\chi$ is shared equally among the products of the reaction. However when the highly quantum non-linear regime $\chi_{1}\gg1$ is entered, $\chi^{\prime}\to \chi_{1}$ and $\chi_{\pm}\to 1$ and the most probable scenario for a highly-relativistic fermion seed that creates a pair is that it remains highly-relativistic and can seed further generations of pairs in a field-free cascade (this point has recently been explored in \cite{king13a}).
\newline

\begin{figure}[!h]
\centering\noindent
  \includegraphics[draft=false, width=8.6cm]{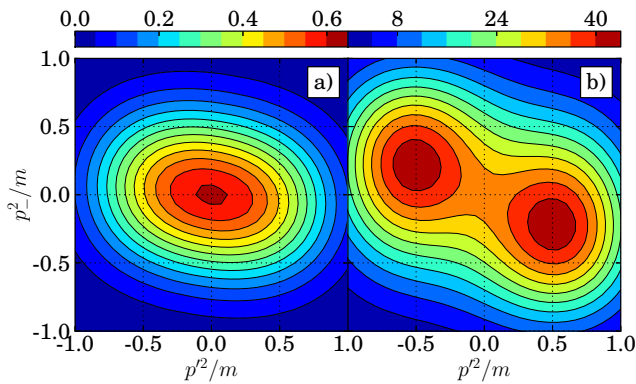}
\caption{The scaled differential rate for the two-step process $10^{8}\partial^{2}\mathcal{I}^{(2)}/\partial p^{\prime y}\partial p_{-}^{y}$ for incoming fermion quantum non-linearity parameter a) $\chi_{1}=1$ and b) $\chi_{1}=10$. For higher $\chi_{1}$, the electrons are split into two ever-distincter momentum regions.} \label{fig:p2yp3yI2}
\end{figure}

By plotting the differential rate in the $2$-momentum component of the outgoing electrons $\partial^{2} \mathcal{I}^{(2)}/ \partial p^{\prime 2}\partial p_{-}^{2}$ in \figref{fig:p2yp3yI2} it can be seen that at high $\chi_{1}$, a beam of electrons colliding head-on with the external field wavevector is split into two connected phase-space regions, denoted by the two lobes in \figrefb{fig:p2yp3yI2}. Using transverse momentum conservation $p_{1}^{2}-p^{\prime 2}-p_{-}^{2}-p_{+}^{2}=0$, it can be shown that the distribution in electron and positron momenta is similarly split, with the line of symmetry along $p_{+}^{2}=-p_{-}^{2}$. The width of the distribution in $p^{\prime 2}$ and $p_{-}^{2}$ can be estimated using the result that bremsstrahlung from an electron is emitted in a cone of radius $\approx 1/\gamma$ \cite{jackson99}. Assuming $\gamma \gg 1$, and that the incoming electron collides head-on with the external field wavevector, the magnitude of the transverse co-ordinate $p_{2}^{\perp}$ over the $3$-
co-ordinate of initial electron momentum must be approximately equal to this angle, i.e. $p^{\prime 2}/p_{1}^{3} \approx 1/\gamma$. Using the approximation $\chi \approx 2\gamma \chi_{E}$, it then follows that $p^{\prime 2}$ and hence $p_{-}^{2}$ must be approximately of the order of unity.

\begin{figure}[!h]
\centering\noindent
  \includegraphics[draft=false, width=5.5cm]{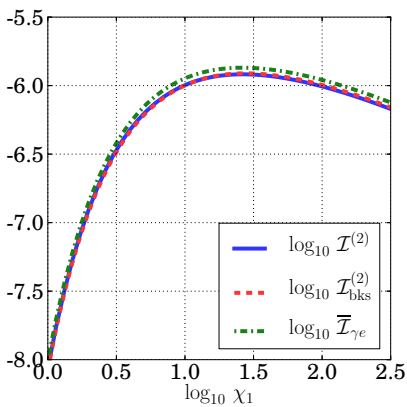}\hfill
\caption{The two-step rate $\mathcal{I}^{(2)}$ is compared to an approximation from the literature $\mathcal{I}_{\trm{bks}}^{(2)}$ and the product approximation of an integration over the lightfront photon momentum using unpolarised sub-processes $\overline{\mathcal{I}}_{\gamma e}$.} 
\label{fig:R22tot}
\end{figure}

The dependency of the total two-step probability on $\chi_{1}$ is plotted in \figref{fig:R22tot}, in which $\mathcal{I}^{(2)}$ is compared to the product approach using unpolarised tree-level differential rates 
\bea
\overline{\mathcal{I}}_{\gamma e} = \int_{0}^{\chi_{1}}\!\!d\chi_{k}~\overline{\mathcal{I}}_{e}\,\frac{\partial\,\overline{\mathcal{I}}_{\gamma}}{\partial\chi_{k}}
\eea
where $\overline{\mathcal{I}}_{j}=(\mathcal{I}_{j,1}+\mathcal{I}_{j,2})/2$ for $j\in\{\gamma,e\}$.
The relative difference between the total unpolarised and polarised rate was found to be around $\approx 12\%$ for $1<\chi_{1}<10^{3}$. In addition, the dashed line in \figref{fig:R22tot} plots an approximate formula derived by Baier, Katkov and Strakhovenko in \cite{baier72}, adapted to the current notation as:
\bea
\mathcal{I}_{\trm{bks}}^{(2)} = \frac{3\alpha^{2}}{16\chi_{1}}\log\left(1+\frac{\chi_{1}}{12}\right)\,\mbox{e}^{-\frac{16}{3\chi_{1}}}(1+0.56\chi_{1}+0.13\chi_{1}^{2})^{\frac{1}{6}}\nonumber, \\
\eea
which was found to agree with $\mathcal{I}^{(2)}$ to within around $4\%$ for $1<\chi_{1}<10^{3}$.

\section{One-step process}
$P^{(1)}$ was evaluated as a five-dimensional numerical integral of the form
\bea
P^{(1)} &=& \xi\Delta\varphi\,\mathcal{I}^{(1)} \label{eqn:P1}  \\
\mathcal{I}^{(1)} &=& \frac{\alpha^{2}}{\pi\chi_{1}}\int\!\! \frac{d\chi_{2}d\chi_{3}dp_{2}^{2}dp_{3}^{2}dv\,\theta(\chi_{1}-\chi_{2}-\chi_{3})}{(\chi_{1}-\chi_{2})^{2}}\frac{\mathcal{B}^{(1)}}{v^{2}}, \nonumber
\eea
where
\bea
\widetilde{\mathcal{B}}^{(1)}(v)&=& |\mathcal{A}^{(1)}(v)-\mathcal{A}^{(1)}(0)|^{2} + \nonumber \\
&& 2\,\trm{Re}\left[\mathcal{A}^{(1)}(0)\left(\mathcal{A}^{(1)\,\ast}(v)-\mathcal{A}^{(1)\,\ast}(0)\right)\right],
\eea
where $\mathcal{B}^{(1)} = \mathcal{B}^{(1)}(v)=\widetilde{\mathcal{B}}^{(1)}(v) + \widetilde{\mathcal{B}}^{(1)}(-v)$ and $\mathcal{A}^{(1)}(v)=\mathcal{A}^{(1)}\left(v,\chi_{1},\chi_{2}, \chi_{3},p_{2}^{2},p_{3}^{2}\right)$ are functions containing products of Airy functions depending on the combination $(p_{2,3}^{2})^{2}+\nu_{2,3}$ with:
\begin{gather}
2^{2/3}\nu_{2,3}(v) = \mu_{2,3}^{2/3} \pm \frac{v}{\chi_{1}\mu_{2,3}^{1/3}} \label{eqn:nus}
\end{gather}
where we note $\nu_{j}(v=0) = (\mu_{j}/2)^{2/3}$, and the integral in $v = 2\chi_{1}\vkap^{0}r/m\chi_{E}$ is between $0$ and $\infty$. One can relate the more complicated function $\mathcal{A}^{(1)}(v)$ to the analytical expression for the two-step process integrand $\mathcal{A}^{(2)}$ by recognising:
\bea
\int\!\!dp_{2}^{2}\,dp_{3}^{2}\,\mathcal{A}^{(1)}(0,\chi_{1},\chi_{2}, \chi_{3},p_{2}^{2},p_{3}^{2}) = \mathcal{A}^{(2)}(\chi_{1},\chi_{2}, \chi_{3}),\nonumber\\ \label{eqn:A1A2}
\eea
where $\mathcal{A}^{(2)}$ was given in \eqnref{eqn:P22}. Numerical integration in the left-hand side of \eqnref{eqn:A1A2} then served as a partial check of $\mathcal{A}^{(1)}(v)$.
\newline

\begin{figure}[!h]
\centering\noindent
  \includegraphics[draft=false, width=8.6cm]{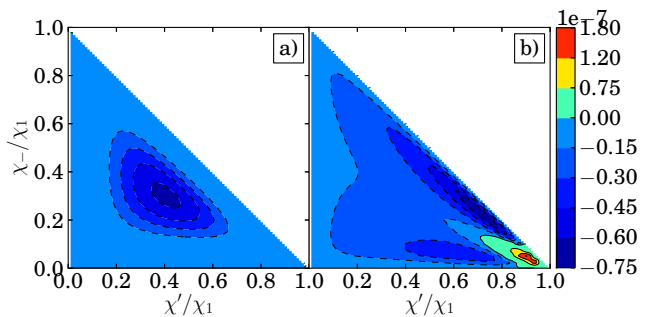}\hfill
\caption{A plot of the differential rate of the one-step process in $\chi^{\prime}$ and $\chi_{-}$ for a) $\chi_{1}=1$ and b) $\chi_{1}=10$. The one-step rate is entirely negative in a).} \label{fig:chi2chi3plotI1}
\end{figure}

As noted in the derivation, the one-step term can be written as the sum of a purely one-step and a one-step-two-step interference term: $\mathcal{I}^{(1)}=\mathcal{I}^{(1)}_{\trm{d}} + \mathcal{I}^{(1)}_{\times}$, where it was seen that $\Iox<0$. To investigate this point, 
we plot in \figref{fig:chi2chi3plotI1} the differential rate $\partial^{2} \mathcal{I}^{(1)}/\partial \chi^{\prime} \partial \chi_{-}$, in which one can clearly observe negative regions. As long as the total differential rate remains positive, this negativity originates from the artificial splitting into one- and two-step terms. The one-step process in these momentum regions is not directly detectable but rather serves to suppress the overall rate. As $\chi_{1}$ increases, the suppression is reduced and the one-step probability becomes positive.
\newline

\begin{figure}[!h]
\centering\noindent
  \includegraphics[draft=false, width=8.6cm]{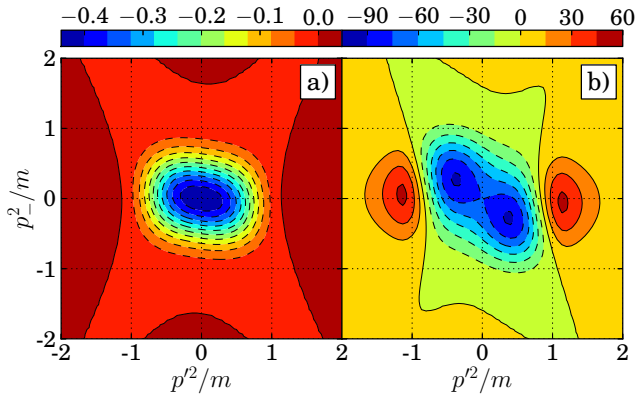}\hfill
\caption{A plot of the differential rate of the one-step process in the $2$-components of the outgoing electrons, multiplied by $10^{8}$ when a) $\chi_{1}=1$ and b) $\chi_{1}=10$. }\label{fig:p2yp3yplotI1}
\end{figure}

To investigate the different dynamics of the one-step process, in \figref{fig:p2yp3yplotI1} we plot the differential rate $\partial^{2} \mathcal{I}^{(1)}/\partial p^{\prime 2}\partial p_{-}^{2}$, in which one can clearly observe that the regions of maximum amplitude are negative. Upon comparison with the two-step differential rate in \figref{fig:p2yp3yI2}, we found that the purely one-step term $\Iod$ splits into two lobes first at much higher $\chi_{1}$ and the line of symmetry in the distribution was at a reduced angle. We consider whether the transverse momentum distribution can be used to measure the one-step process in the \sxnref{sxn:totalrateA}.
\newline

\begin{figure}[!h]
\centering\noindent
  \includegraphics[draft=false, width=7cm]{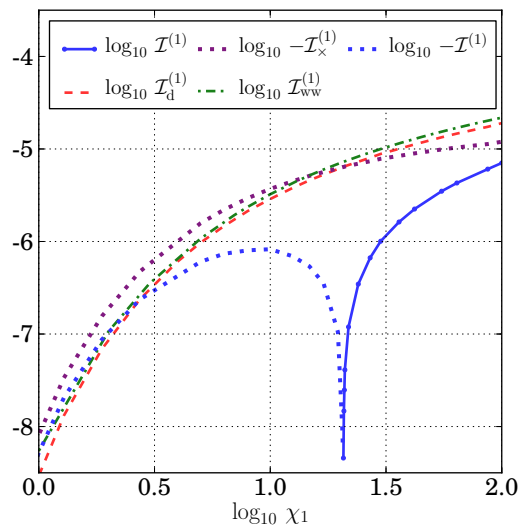}\hfill
\caption{Various parts of the one-step probability. Dotted lines represent negative contributions (suppression), and for $\chi_{1}\gtrsim20$, the total one-step probability becomes positive (solid line). The purely direct term $\mathcal{I}_{\trm{d}}^{(1)}$ is also compared with the Weizs\"acker-Williams approximation $\mathcal{I}_{\trm{ww}}^{(1)}$.} \label{fig:I1tot}
\end{figure}

The total probability of the one-step process is plotted in \figref{fig:I1tot}. We note that the total one-step rate becomes positive for $\chi_{1}\gtrsim 20$. This is a consequence of the positive purely one-step rate, which increases with $\chi_{1}$ becoming dominant over the negative cross term between one-step and two-step processes, in which the two-step rate decreases for $\chi_{1}\geq 20$. One could consider the following intuitive reasoning as to why the off-shell part increases with $\chi_{1}$, in contrast to the on-shell part. From the bandwidth theorem applied to the uncertainty relation, $\Delta \mathcal{E} \Delta T \sim \hbar$, the spread of virtual energies available $\Delta\mathcal{E}$ is larger when the interaction time $\Delta T$ is shorter, which here corresponds to the time taken to traverse the formation length $L_{\ast}$. Since $\Delta T\propto 1/u_{1}$, where $u_{1}$ is the velocity of the incoming electron, which increases with $\chi_{1}$, the virtual photons can therefore more easily fulfil energy-momentum conservation at both vertices. The two-step process on the other hand, does not benefit from this scaling.  We discuss the consequences 
of this in \sxnref{sxn:totalrate}. \\

From \figref{fig:I1tot} we note that the maximum suppression of the two-step rate due to a negative one-step rate occurs at $\chi_{1}\approx 10$. Since the cross-term is essentially the overlap between the one-step and two-step process, and since this grows with $\chi_{1}$, the suppression does not appear to be due to the increasing overlap of the two decay channels as in e.g. the LPM (Landau-Pomeranchuk-Migdal) effect (recently studied in the combination of laser and atomic fields \cite{dipiazza12b}). What can be noticed is that the probability for the one-step process becomes only positive when, over a formation length $L_{\ast}$, regardless of the extent of the field, the purely one-step process becomes more probable than the two-step process.

\begin{figure}[!h]
\centering\noindent
  \includegraphics[draft=false, width=5cm]{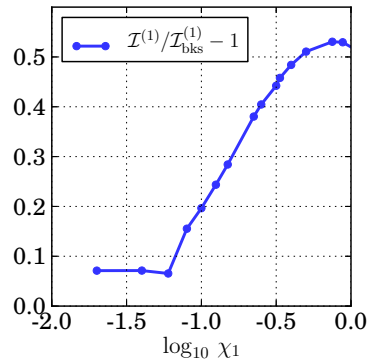}\hfill
\caption{The agreement between the one-step rate and the asymptotic expression $\mathcal{I}_{\trm{bks}}^{(1)}$ from \cite{baier72} for $\chi_{1}\ll 1$.} \label{fig:I1asy}
\end{figure}

The total rate for the one-step process can also be verified in part by using existing asymptotic approximations in the literature, for example from \cite{baier72} one has:
\begin{equation}
\mathcal{I}^{(1)}_{\trm{bks}} \sim \frac{-\alpha^{2}}{32}\sqrt\frac{2\chi_{1}}{3\pi}\mbox{e}^{-\frac{16}{3\chi_{1}}}, \qquad \chi_{1} \ll 1\\
\end{equation}
the error for which remains less than $20\%$ for $\chi_{1} \lesssim 0.1$, as plotted in \figref{fig:I1asy}. 

A test that was used to approximate the one-step background in E-144 \cite{burke97, bamber99}, is the Weizs\"acker-Williams (WW) approximation $P_{\trm{ww}}$ \cite{weizsaecker34,*williams34, jackson99}, which is applicable to the purely virtual part, $P^{(1)}_{\trm{d}}$. This approximation substitutes the virtual photon spectrum of a charged seed particle with an equivalent real photon spectrum (by neglecting non-transverse polarisation components) and then assumes the individual real photon frequency components can be summed incoherently with the photon-seeded pair-creation probability. One can achieve a similar result by performing the spin trace over just the transverse spacetime indices, followed by a limiting procedure $k^{2}\to 0$ in the resulting quantum amplitudes \cite{olsen79}. This should be a good approximation to the trident process, the closer the intermediate virtual photon is to being real, i.e. the smaller $k^{2}$ becomes with respect to the electron mass. We note:
\bea
k^{2} &=& (p_{2}-p_{1})^{2} - 2r\vkap (p_{2}-p_{1}),
\eea
therefore the closer the scattered electron momentum is to the original one, i.e. the smaller the electron recoil, the smaller $k^{2}$ is, independent of $r$. For the two-step case, the integration of the product approximation shows \cite{king13a} that in general $\chi_{k} \ll \chi_{1}$ and $\chi_{2}\approx \chi_{1}$ when $\chi_{1}\gg 1$. It then follows that when $\chi_{1}\gg 1$, the Weizs\"acker-Williams method can be used to approximate to the directly virtual part of the probability, $P^{(1)}_{\trm{d}}$. One can modify the Weizs\"acker-Williams approximation for pair-creation via bremsstrahlung \cite{heitler60} to the present case 
\bea
P^{(1)}_{\trm{ww}}(\chi_{1})\!=\!\frac{2\alpha}{\pi}\!\int_{0}^{\chi_{1}}\!\frac{d\chi_{k}}{\chi_{k}}\!\left[\ln \left(\frac{\chi_{1}}{\chi_{k}}\right)\!-\!C\right]\!P_{e}(\chi_{k}),
\eea
where $C=\gamma_{E}+1/2-\ln2\approx 0.384$, $\gamma_{E}=0.57721\ldots$ is the Euler constant. The comparison is made in \figref{fig:I1tot} where we have taken $P_{e}=\xi\Delta\varphi\, \overline{\mathcal{I}}_{e}$ and $P^{(1)}_{\trm{ww}}=\xi\Delta\varphi\,\mathcal{I}^{(1)}_{\trm{ww}}$. For values of $\chi_{1}\lesssim2$, the WW approximation becomes quickly worse than $10\%$, seeming to support the decomposition of the one-step process into purely-virtual and cross-term parts (although the error remains of the order of the polarisation error in the 2-step process, of around $10\%$). However, despite the accuracy when compared to the purely virtual one-step process, the WW approach does not take into account the cross-term between one- and two-step processes and therefore even fails to indicate the region of measurability of the one-step process or its suppressive effect on the total rate. One might conjecture that the WW rate could be useful when $\chi_{1}$ increases above $100$, but we have not investigated this here as doubt has been cast on whether the perturbative expansion in final particles is valid at such large value of $\chi_{1}$ \cite{narozhny80}.
\newline

\section{Total trident process rate} \label{sxn:totalrate}
\begin{figure}[!h]
\centering\noindent
  \includegraphics[draft=false, width=8.6cm]{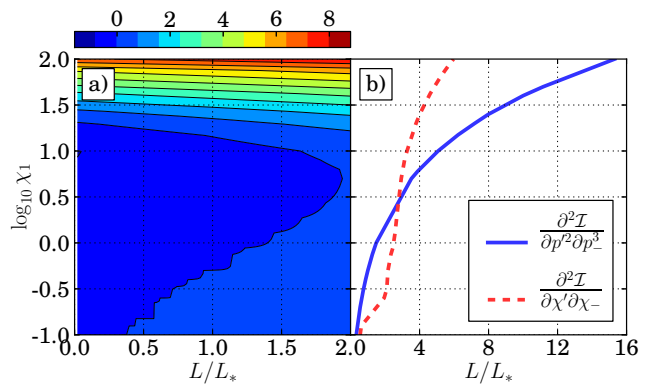}\hfill
\caption{Plot a) is the sum of the total two- and one-step rates (multiplied by $10^{6}$) as a function of incoming non-linear quantum parameter $\chi_{1}$ and external field dimension $L/L_{\ast}$. Plot b) is the maximum value of $L/L_{\ast}$ such that the differential rate in $p^{\prime 2}$ and $p_{-}^{2}$ (solid line) and $\chi^{\prime}$ and $\chi_{-}$ (dashed line) remain positive.} \label{fig:Itot}
\end{figure}

Comparison of the one- and two-step processes depends not only $\chi_{1}$ but also on the extent of the external field, and will allow for an analysis of how small the external field can be taken without substantially violating the assumptions in the derivation. In previous sections, it was seen that for small $\chi_{1}$, the total one-step process becomes negative. Since the total rate must remain positive, this allows for a condition on the minimum allowable dimension of external field. To make the discussion more transparent let us rewrite the phase factors in terms of formation lengths recalling $\xi\Delta\varphi = L/L_{\ast}$, $L_{\ast}=\lambdabar/\chi_{E}$, so that the total probability is of the form:
\bea
P = \frac{1}{2}\left(\frac{L}{L_{\ast}}\right)^{2}\mathcal{I}^{(2)} + \frac{L}{L_{\ast}}~\mathcal{I}^{(1)} + \mathcal{I}^{(0)},
\eea
where $\mathcal{I}^{(0)}$ corresponds to the neglected interference between exchange terms. By regarding $L/L_{\ast}$ as a separate variable, we plot the dependency of the calculated terms in $P$ on both $\chi_{1}$ and $L/L_{\ast}$ in \figref{fig:Itot}. From the plot of the total rate in \figrefa{fig:Itot}, it seems that $L/L_{\ast}<2$ is required before the total rate becomes negative, and this is most critical at around $\chi_{1}\approx 10$. One might argue that since $L/L_{\ast}<2$, the $\mathcal{I}^{(0)}$ term must be taken into account and possibly cancels out this negativity. However, in \figrefb{fig:Itot} we plot the maximum ratio of $L/L_{\ast}$ required such that also the differential rate of the calculated terms in $p^{\prime 2}$ and $p_{-}^{2}$ (solid line) and in $\chi^{\prime}$ and $\chi_{-}$ (dashed line) remains positive, which is seen to grow with $\chi_{1}$ into the $L/L_{\ast}>2$ region ($L/L_{\ast}\approx 60$ was found for $\chi_{1}=10^{3}$, although it is unknown whether the current 
method is at all applicable at such high $\chi$-factors \cite{narozhny80}). Therefore the $\mathcal{I}^{(0)}$ term is unlikely to be fundamental to the discussion. 

The origin of this negativity is most likely the assumptions made in deriving the constant crossed field rates. Most notably, the infinite integral over the external-field phase performed to generate the characteristic Airy functions must be modified to include the finite duration. These results imply a useful additional constraint on the validity of the constant crossed field approximation, namely
\bea
\xi \Delta \varphi \gg 1.
\eea
This is independent of the condition $\xi\gg1$, which is required to justify the limit on the external-field frequency $\varkappa^{0}\to 0$ in the expression for a general background because $\xi \Delta \varphi$ is independent of $\varkappa^{0}$.\\


\subsection{Measurability} \label{sxn:totalrateA}
\begin{figure}[!h]
\centering\noindent
  \includegraphics[draft=false, width=4.25cm]{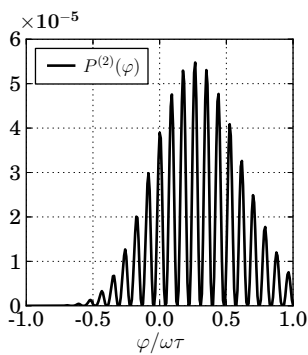}\hfill
\caption{The maximum of the probability of the two-step process using the locally constant crossed field approximation is shifted from the maximum of the external field at $\varphi=0$.} \label{fig:lcfapic}
\end{figure}

\begin{figure}[!h]
\centering\noindent
\includegraphics[draft=false, width=8.6cm]{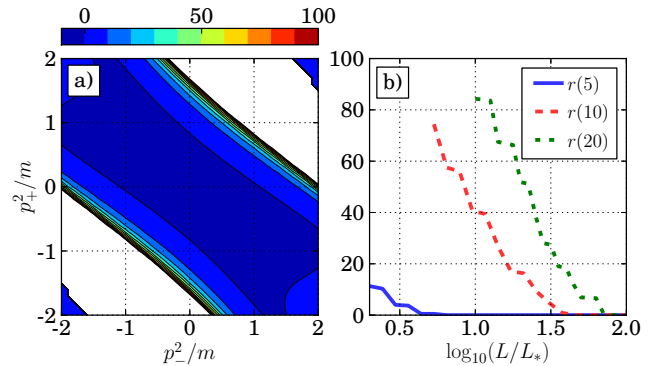}
\caption{In plot a) for $\chi_{1}=10$ is the ratio of the one-step to the two-step process, where in the empty regions, the ratio becomes much larger than the maximum on the colour scale. In plot b), the percentage $r(\chi_{1})$ of the measurable one-step signal to the total signal in the $p_{-}^{2}$--$\,p_{+}^{2}$ plane, with increasing $L/L_{\ast}$, is plotted. } \label{fig:p3yp4yI}
\end{figure}

Here we are discussing the measurability of electron-seeded pair creation in the collision of a laser and electron beam in the non-perturbative and highly non-linear regime ($\xi\gg1$) in contrast to the weakly nonlinear regime ($\xi\lesssim 1$) discussed in \cite{hu10} in a curtailed plane-wave background. The two-step process is measurable by taking similar parameters to the SLAC E-144 experiment, but increasing the intensity of the laser used from $1.3\times10^{18}~\trm{Wcm}^{-2}$ to $10^{22}~\trm{Wcm}^{-2}$ (the frequency $\omega=527\,\trm{nm}$) and reducing the pulse duration from $\tau=1.6~\trm{ps}$ to $\tau=10~\trm{fs}$ allowing the energy and number of initial electrons to be reduced to  $10^{9}$, $2~\trm{GeV}$ electrons. Assuming the slight angle between the laser and particle beam required in any experimental set-up makes only a minor difference to our analysis for a head-on collision (in E-144 the angle was $\pi/10$ radians and $\chi_{1}(\theta)\approx \chi_{1}(0)(1-\theta^{2}/2)$ for collision angle $\theta\ll1$), $\chi_{1} = 1$, $\xi = 30$, $L/L_{\ast}=1100$, for a constant external field, the initial beam of electrons would create of the order of $5\times10^{6}$ pairs. This estimate can be improved by using the  locally constant crossed field approximation:
\bea
P^{(2)}&=&\int_{-\!\infty}^{\infty}\!\!d\varphi_{x} \int_{0}^{\chi_{1}}\!\!d\chi_{k}\,\frac{\partial P_{\gamma}[\chi_{1}(\varphi_{x}), \chi_{k}(\varphi_{x})]}{\partial \chi_{k}} \nonumber \\
&& \qquad\qquad \int_{-\!\infty}^{\varphi_{x}}d\varphi_{y}\,P_{e}[\chi_{k}(\varphi_{y})],
\eea
for a $500\,\trm{TW}$ laser pulse of focal width $2\,\mu\trm{m}$ modelled by $E(\varphi)=E_{0}\,\mbox{e}^{-(\varphi/\varphi_{0})^{2}}\!\cos\varphi$, where $\varphi_{0}=\omega\tau$, ($\xi\leq30$, $\chi\leq 1$), the probability of which is plotted in \figref{fig:lcfapic}, which then predicts of the order of $10^{5}$ pairs.
However, we stress that Compton scattering and focusing effects have been neglected, which would act to reduce this estimate. The one-step process in this regime seems much more difficult to separate in an experiment. One possibility is to measure the transverse momenta of final state positrons (the electron momentum distribution will likely be washed out by double-photon Compton scattering of the initial electron). Due to the slightly wider momentum spectrum in the $2$-direction for positrons created via the one-step process compared to the two-step process, plotted in \figrefa{fig:p3yp4yI}, with a judicious momentum ``cut'', the effect of the one-step process could be measured. In \figrefb{fig:p3yp4yI} the ratio $r(\chi_{1})$ of one-step to two-step created positrons outside of the area $|p_{\pm}^{2}|<m$, in regions of the detector where the former is at least $50\%$ the latter, is plotted as a function of the field dimension $L/L_{\ast}$. The one-step process is then most measurable for a head-on collision when the $\chi$-factor is increased but $\chi_{E}$ held relatively low. A beam of $10^{9}$, $250~\trm{GeV}$ electrons combined with a single-cycle $10^{21}~\trm{Wcm}^{-2}$ laser pulse ($\chi_{1}=40$,  $\xi=10$, $L/L_{\ast}\approx 60$) would produce of the order of $10^{4}$ positrons via the one-step process in these measurable regions from initial seed electrons. Another possibility not explored here, is that only in those parts of the pulse for which $\chi_{1}\gtrsim20$ lead to one-step pair creation, and so if the created positrons could be ``streaked'' \cite{itatani02, kitzler02}, one could perhaps utilise this well-defined phase-space region to better exclude the two-step background. Although further Compton-scattering of the positrons could take place, this three- rather than a two- vertex dressed process is less probable. 

\section{Discussion} 
A fundamental difference between the two- and one- step processes is that in the two-step process, the real photon can propagate for an arbitrarily long phase length before decaying into a pair, whereas in the one-step process, the photon's range is limited by the uncertainty relation. So when the sub-processes are not very probable, for example at low $\chi$, one would expect the two-step process to dominate. If the range of the real photon were curtailed to a length of the order of the formation length, the length associated with the one-step process, this advantage of the two-step process would be lost and the two processes should be comparable, as we indeed find. On the other hand, considering the uncertainty relation in energy and time, the shorter the interaction duration, the larger the bandwidth of frequencies available to fulfill energy-momentum conservation at the vertices of one-step pair creation. Therefore, for electrons incident with a high $\chi$ parameter, the probability of the one-step 
process is enhanced compared to the two-step version, leading to its dominance for external field dimensions of the order of the formation length and large $\chi$.\\

An important point is the overlap of the two production channels. Assuming that the constant crossed field is still valid ($\xi \Delta \varphi\gg1$), instead of considering the suppression of the two-step through the one-step process, one could instead consider this phenomenon the other way round. The one-step process is essentially suppressed by the more probable two-step process at low $\chi$, but when $\chi$ is large enough that the one-step channel becomes more probable over its formation length, the one-step process probability becomes positive and emerges as a separate channel on its own. Moreover, this threshold is independent of the extent of the external field and so this phenomenon is likely present in more complicated backgrounds as well.\\

Although we have only calculated a two-step process, the question presents itself whether the constant crossed field approximation can be used to model long chains of processes as is typically used in computer codes. In this case for $n$ vertices, assuming the Poisson-like dependence on the external field's spatial extent (and duration) continues to higher orders, one could have a probability of the form:
\bea
P = \frac{1}{n!}\left(\frac{L}{L_{\ast}}\right)^{n}~\mathcal{I}^{(n)} +\ldots + \frac{L}{L_{\ast}}~\mathcal{I}^{(1)} + \mathcal{I}^{(0)},
\eea
where $\mathcal{I}^{(j)}$ represents a process with $j-1$ on-shell propagators. Assuming the dynamical factors $\mathcal{I}^{(j)}$ are of the same order, the potentially largest negative contribution arises from the $\mathcal{I}^{(n)}$--$\mathcal{I}^{(n-1)}$ cross-term, which acquires an extra factor $n$. Therefore one might surmise that the constant-crossed-field approximation is also valid for an $n$-vertex process as long as $\xi \Delta \varphi\gg n$, although calculation of higher order processes would be necessary in order to validate such speculation. \\

The two-step process was measured in the weakly non-linear (multi-photon) regime ($\xi=0.3$) in the E-144 experiment. We calculated the approximate number of pairs created in the highly non-linear and non-perturbative regime ($\xi\gg1$, $\chi>1$). It was found that if the laser intensity could be updated to higher, currently available intensities ($2\times10^{22}~\trm{Wcm}^{-2}$ was already achieved in 2008 \cite{yanovsky08}), the particle beam can be allowed to be even less relativistic and the two-step process measurable. More difficult is separating the one-step mechanism for which it is crucial that the field extent is not much larger than the formation length otherwise the background from the two-step channel is too large. However, to utilise the slightly different transverse positron momentum distribution, the $\chi$ factor must remain high $\chi\gtrsim 20$. Here around $10^{4}$ positrons were predicted to be produced in ``measurable'' regions for a single-cycle $10^{21}~\trm{Wcm}^{-2}$ pulse but a $250~\trm{GeV}$ electron beam. These results for the $\xi\gg1$ ``quasi-static'' or as often referred to ``tunnelling'' regime can be contrasted with the analysis in \cite{hu10}, in which it was shown that in the $\xi\ll1$ ``multi-photon'' regime, in which probabilities involving $N$ photons are proportional to $\xi^{2N}$, if the frequency of the laser pulse can be made high enough in the rest frame of the seed particles, the one-step process (leading order $N=2$) can be orders of magnitude larger than the two-step process (leading order $N=3$).\\


\section{Conclusion} 
The trident process in a constant crossed field must be considered in its entirety, being separable into two- and one- step processes rather than real and virtual parts, which were both seen to contribute to the one-step process. The two-step process was found to agree exactly with an integration of the average of polarised tree-level processes over lightfront momenta. The one-step process was found to be effectively suppressed for $\chi_{1}\lesssim20$ due to the larger probability of the two-step process over the formation length. In the highly non-linear and non-perturbative regime ($\xi\gg1$, $\chi>1$), the two-step process was shown to be measurable for electron beams even less energetic than in the E-144 experiment, as long as the intensity of the laser is updated to around $10^{22}~\trm{Wcm}^{-2}$. For field dimensions not orders of magnitude larger than the formation length, it was shown that the one-step process could in principle be separated from the two-step process using the wider positron transverse momentum spectrum when a single-cycle $10^{21}~\trm{Wcm}^{-2}$ pulse collides with a $250~\trm{GeV}$ electron beam.

\section{Acknowledgments}
B. K. would like to acknowledge many stimulating discussions with A. Ilderton and A. Fedotov and conversations with P. B\"ohl. This work was performed at the Arnold Sommerfeld Center for Theoretical Physics and supported by Grant No. DFG, FOR1048, RU633/1-1, by SFB TR18 project B12 and by the Cluster-of-Excellence ``Munich-Centre for Advanced Photonics'' (MAP). Plots were generated with {\tt{Matplotlib}} \cite{matplotlib}.


\appendix
\section{Further detail on derivation}
\subsection{Definitions \label{app:deriv}}
Here we define objects used in the manuscript and further calculation. The Volkov states are \cite{landau4}:
\bea
\psi_{r}(p) &=& \Big[1+\frac{e \sk\sA}{2\varkappa p}\Big] \frac{u_{r}(p)}{\sqrt{2p^{0}V}} \mbox{e}^{iS(p)} \nonumber\\
\psibar_{r}(p) &=& \frac{\overline{u}_{r}(p)}{\sqrt{2p^{0}V}}\Big[1+\frac{e \sA\sk}{2\varkappa p}\Big] \mbox{e}^{-iS(p)}\nonumber\\
\psi^{+}_{r}(p) &=& \Big[1-\frac{e \sk\sA}{2\varkappa p}\Big] \frac{v_{r}(p)}{\sqrt{2p^{0}V}} \mbox{e}^{iS(-p)} \nonumber\\
S(p) &=& -px - \int^{\varphi}_{0} d\phi \,\Big(\frac{e(pA[\phi])}{\vkap p} - \frac{e^{2}A^{2}[\phi]}{2(\vkap p)}\Big),
\label{eqn:Volkov}
\eea
where $\slashed{\varkappa}=\gamma^{\mu}\varkappa_{\mu}$, $\gamma^{\mu}$ are the gamma-matrices, $u_{r}$ ($v_{r}$) are free-electron (-positron) spinors satisfying $\sum_{r=1}^{2}u_{r\rho}(p)\overline{u}_{r\sigma}(p)=(\slashed{p}+m)_{\rho\sigma}/2m$, $\sum_{r=1}^{2}v_{r\rho}(p)\overline{v}_{r\sigma}(p)=(\slashed{p}-m)_{\rho\sigma}/2m$, $\overline{u} = u^{\dagger}\gamma^{0}$ and the remaining symbols are as described in the paper. The photon propagator is:
\bea
G^{\mu\nu}(x-y) = \int \frac{d^{4}k}{(2\pi)^{4}} \frac{4\pi g^{\mu\nu}}{k^{2}+i\eps}\,\mbox{e}^{ik(x-y)}. \label{eqn:photprop}
\eea
The expansion of the vertices in Fourier modes is:
\bea
\int \,\frac{dr}{2\pi} ~\Gamma^{\mu} (r) \mbox{e}^{-ir\varphi}  &=&  ~\psibar_{2}(\varphi)\, \gamma^{\mu}\, \psi_{1}(\varphi)\\
\Gamma^{\mu}(r) &=& \int \!d\varphi~ \mbox{e}^{ir\varphi} ~\psibar_{2}(\varphi)\, \gamma^{\mu}\, \psi_{1}(\varphi)\label{eqn:Gam1}\\
\int \,\frac{ds}{2\pi} ~\Delta^{\mu} (s) \mbox{e}^{-is\varphi}  &=&  ~\psibar_{3}(\varphi)\, \gamma^{\mu}\, \psi^{+}_{4}(\varphi)\\
\Delta^{\mu}(s) &=& \int d\varphi~ \mbox{e}^{is\varphi} ~\psibar_{3}(\varphi)\, \gamma^{\mu}\, \psi^{+}_{4}(\varphi)\label{eqn:Del1},
\eea
where we have used the shorthand $\psi_{i}$ = $\psi(p_{i})$ with spinor indices suppressed and $\psi_{j}(\varphi)$ are the Volkov states with Fourier terms $\mbox{e}^{\pm i p_{j}x}$ removed.

\subsection{Derivation of rate expression}
Beginning from the expression for the scattering matrix:
\bea
S_{fi} &=& \alpha\!\!\int\! d^{4}x \,d^4y \,\, \psibar_{2}(x)\gamma^{\mu}\psi_{1}(x)G_{\mu\nu}(x-y)\psibar_{3}(y)\gamma^{\nu}\psi^{+}_{4}(y) \nonumber \\ && \qquad - \,(p_{2}\leftrightarrow p_{3}), \label{eqn:Sfi1b}\\
 &=& \overrightarrow{S_{fi}} - \overleftarrow{S_{fi}}.
\eea
Using the definitions in \eqnreft{eqn:Volkov}{eqn:Del1}, we can rewrite \eqnref{eqn:Sfi1b} as:
\bea
\overrightarrow{S_{fi}} &=& \frac{\alpha}{\pi}\!\!\int\!\!d^{4}x\, d^{4} y \,\frac{d^{4} k}{(2\pi)^{4}} \,dr\,ds ~\mbox{e}^{ix\Pi_{\Gamma}+ iy\Pi_{\Delta}}\frac{\Gamma^{\mu}(r)\,\Delta_{\mu}(s)}{k^{2}+i\varepsilon},\nonumber \\ \label{eqn:Sfi2}
\eea
where $\Pi_{\Gamma} = k-\delta p_{\Gamma}$, $\delta p_{\Gamma} = p_{1}+r\vkap-p_{2}$ and $\Pi_{\Delta} = -k-\delta p_{\Delta}$, $\delta p_{\Delta} = s\vkap-p_{3}-p_{4}$. Performing the integration in \eqnref{eqn:Sfi2} over $x$ and $y$ gives:
\bea
\overrightarrow{S_{fi}} &=& \frac{(2\pi)^{4}\alpha}{\pi}\!\!\int\!\!d^{4} k \,dr\,ds~\delta(k-\delta p_{\Gamma})\nonumber\\
&& \qquad\qquad\times\delta(k+\delta p_{\Delta})\frac{\Gamma^{\mu}(r)\,\Delta_{\mu}(s)}{k^{2}+i\varepsilon},\nonumber \\ \label{eqn:Sfi2c}
\eea
and over $k$ gives:
\bea
\overrightarrow{S_{fi}} &=& \frac{(2\pi)^{4}\alpha}{\pi}\!\! \int\!\!dr\,ds ~\delta^{(4)}(\Delta p -(r+s)\vkap)\nonumber \\
&&\qquad\qquad\times\frac{\Gamma^{\mu}(r)\,\Delta_{\mu}(s)}{k'^{2}+i\varepsilon}\Big|_{k'=k'_{\ast}}, \nonumber \\\label{eqn:Sfi3b}
\eea
where $k'_{\ast} = \delta p+r\vkap$, $\delta p = p_{1}-p_{2}$ and $\Delta p = p_{2}+p_{3}+p_{4}-p_{1}$. We notice:
\bea
\frac{1}{k'^{2}+i\varepsilon}\Big|_{k'=\delta p+r\vkap} = \frac{1}{(\delta p)^{2}+2r\vkap \delta p + i\eps} = \frac{(2\vkap \delta p)^{-1}}{r-r_{\ast} + i\eps}, \nonumber \\
\eea
where we have defined $r_{\ast} = -(\delta p)^{2}/(2\vkap \delta p)$. With a co-ordinate transformation $r\to r+r_{\ast}$ we have:
\bea
\overrightarrow{S_{fi}} &=& \frac{(2\pi)^{3}\alpha}{\vkap \delta p}\!\! \int\!\!\frac{dr\,ds}{r+i\eps} ~\delta^{(4)}(\Delta p -(r+r_{\ast}+s)\vkap)\nonumber \\ && \qquad\qquad\qquad\qquad \Gamma^{\mu}(r+r_{\ast}) \,\Delta_{\mu}(s), \label{eqn:Sfi4b}
\eea
In order to evaluate the delta functions, we switch at this point to lightcone co-ordinates. For spatial co-ordinates we define $x^{\pm} = (x^{0} \pm x^{3})$, $x_{\perp} = (x_{1}, x_{2})$, whereas for momenta, $p^{\pm} = (p^{0}\pm p^{3})/2$, $p_{\perp} = (p_{1},p_{2})$. We also define a  co-ordinate system and specify a constant crossed field $\varkappa = \varkappa^{0}(1,0,0,1)$, $A^{\mu}(\varphi)=a^{\mu}\varphi$, $a^{\mu} = (E/\varkappa^{0})(0,1,0,0)$, $\varkappa a = \varkappa^{2} = 0$, so that $\vkap x = \vkap^{0} (x^{0}-x^{3}) = \vkap^{+}x^{-}$. 

In forming the probability, we must square the scattering matrix. Let us concentrate on $|\overrightarrow{S_{fi}}|^{2}$ as the steps for other contributions are similar. When \eqnref{eqn:Sfi3b} is mod-squared, one has, for some function $f=f(r,s,r',s')\in\mathbb{C}^{\infty}$:
\begin{widetext}
\bea
|S|^{2} &=&\int dr\,dr'\,ds\,ds'~f\,\delta^{(4)}[\Delta p - (r + s)\vkap]~\delta^{(4)}[\Delta p - (r' + s')\vkap] \label{eqn:form1eqS}\\
 &=& \int dr\, dr'\,ds\,ds'~f\,\delta^{(4)}[\Delta p - (r + s)\vkap]~\delta^{(4)}[
(r+s- (r'+s'))\vkap] \\
&=& \int dr\, dr'\,ds\,ds'~f\,\delta^{(4)}[\Delta p - (r + s)\vkap]~\frac{\delta^{(4)}[
(r+s- (r'+s'))\vkap]}{\delta(r+s- (r'+s'))}\delta(r+s- (r'+s'))\\
 &=& \int dr\, dr'\,ds\,ds'~f\,\delta^{(4)}[\Delta p - (r + s)\vkap]\frac{VT}{(2\pi)^{3}\Delta\varphi_{+}} \delta(r+s- (r'+s'))\\
 &=& \frac{VT\delta^{(2)}(\Delta p^{\perp})\delta(\Delta p^{-})}{(2\pi \vkap^{0})^{3}\Delta\varphi_{+}} \int d\tilde{r}\,d\tilde{r}'~f(\tilde{s}=\Delta p-\tilde{r},\tilde{s}'=\Delta p-\tilde{r}'), \label{eqn:form1eqE} \label{eqn:Sfi4c}
\eea
\end{widetext}
where we have defined a formation phase length $\Delta\varphi_{+}$ \cite{ritus85}:
\bea
\delta(r+s-(r'+s'))\Big|_{r+s=r'+s'}&=&\frac{\Delta\varphi_{+}}{2\pi}, \label{eqn:Ldef}
\eea
where
\bea
\delta(x)\big|_{x=0} &=& \int \!\frac{dl}{2\pi}~ \mbox{e}^{ixl}\Big|_{x=0},
\eea
and $\tilde{q}:=\vkap^{0}q$ for $q\in\{r,r',s,s'\}$.

At this point, since we wish to form probabilities and not rates, we invoke the relation $T/p_{1}^{0} = \Delta\varphi_{+}/\varkappa p_{1}$ \cite{ritus85}, so that, combining the arguments leading to \eqnref{eqn:Sfi4b} and \eqnref{eqn:Sfi4c}, we then have:
\bea
\left|\overrightarrow{S_{fi}}\right|^{2} &=& \frac{(2\pi)^{3}\alpha^{2}}{(\vkap \delta p)^{2}} \frac{Vp_{1}^{0}\mathcal{I}(\rightarrow,\rightarrow)}{\vkap^{0}(\vkap p_{1})} \delta^{(2)}(\Delta p^{\perp})\delta(\Delta p^{-}) \nonumber\\
\mathcal{I}(\rightarrow,\rightarrow) &=& \left|\int dr~\frac{\Gamma^{\mu}(r+r_{\ast})\Delta_{\mu}(s_{\ast} -r)}{r+i\eps}\right|^{2}  \label{eqn:Sfi4d}
\eea
where we have defined $s_{\ast} = \Delta p^{+}/\vkap^{0} - r_{\ast}$, which can be shown to be equal to:
\bea
s_{\ast} = \frac{(p_{2}+p_{3}+p_{4})^{2}-m^{2}}{2p_{1}\vkap} -r_{\ast}.
\eea 
We note that in order to evaluate the light-cone co-ordinate delta functions occurring in \eqnref{eqn:Sfi4d} from a Cartesian integral, one can use:
\bea
\int \frac{d^{3}p_{j}}{2p_{j}^{0}} f(p_{j}) = \int \frac{d^{2}p_{j}^{\perp} dp^{-}}{2p^{-}_{j}} \theta(p_{j}^{-})f(p_{j})\Bigg|_{p_{j}^{+}=\frac{(p_{j}^{\perp})^{2}+m^{2}}{4p_{j}^{-}}}, \label{eqn:Sfi4e}
\eea
where $\theta(\cdot)$ is the Heaviside step function.
\newline

The probability $\overrightarrow{P}$, using the expression $\overrightarrow{P} = (1/2)\prod_{j=2}^{4}[V\int d^{3}p_{j}/(2\pi)^{3}] \tr|\overrightarrow{S_{fi}}|^{2}$, is then given by:
\bea
\overrightarrow{P} &=& \frac{\alpha^{2}}{ 2^{6}(\vkap^{0})^{3} (\vkap p_{1})}\prod_{j=2,3} \int \frac{d^{2}p^{\perp}_{j}}{(2\pi)^{3}} \frac{dp^{-}_{j}}{p^{-}_{j}}\frac{\theta(p_{j}^{-})\tr\mathcal{I}(\rightarrow,\rightarrow) \,\big|_{\trm{nn}}}{p_{4}^{-}(p_{1}^{-}-p_{2}^{-})^{2}}, 
\nonumber \\
\eea
where the instruction $\textrm{nn}$ means that all normalisations of the form $(2Vp_{j}^{0})^{-1/2}$ for $j \in \{1,2,3,4\}$ have been removed and the integral in $d^{3}p_{4}$ has already been performed (to account for the degeneracy of outgoing states the total probability $P$ requires an extra factor $1/2$ as explained in the main text).

\subsection{Vertex functions}
We can rewrite the vertex functions \eqnrefs{eqn:Gam1}{eqn:Del1} in a way that allows them to be easily evaluated by separating integrals from trace products. Concentrating first on $\Gamma^{\mu}(r)$:
\bea
\Gamma^{\mu}(r) &=&\!\int\!\!d\vphi \left\{ \frac{\overline{u}_{\sigma_{2}}(p_{2})}{\sqrt{2p_{2}^{0}V}}\Big[1+\frac{e\slashed{A}(\vphi)\slashed{\vkap}}{2\vkap p_{2}} \Big]\gamma^{\mu}\right. \nonumber \\
&&\left.\Big[1+\frac{e\slashed{\vkap}\slashed{A}(\vphi)}{2\vkap p_{1}}\Big]\frac{u_{\sigma_{1}}(p_{1})}{\sqrt{2p_{1}^{0}V}}\mbox{e}^{i ( r\vphi + c_{2} \vphi^{2} + c_{3} \vphi^{3})}\right\}, \label{eqn:Gamma1}
\eea
where we have introduced:
\begin{gather}
c_{2} \defto \frac{e}{2}\Big(\frac{p_{2}a}{\vkap p_{2}} - \frac{p_{1}a}{\vkap p_{1}}\Big); ~~ c_{3} \defto -\frac{e^{2}a^{2}}{6}\Big(\frac{1}{\vkap p_{2}} - \frac{1}{\vkap p_{1}}\Big). \label{eqn:alphabeta}
\end{gather}

Now as $A^{\mu} = a^{\mu}\vphi$, we can rewrite \eqnref{eqn:Gamma1} as:
\bea
\Gamma^{\mu}(r) &=&  \frac{\overline{u}_{\sigma_{2}}(p_{2})}{\sqrt{2p_{2}^{0}V}} \Big[C_{1}\gamma^{\mu}+C_{2}\frac{e}{2}\Big(\frac{\slashed{a}\slashed{\vkap}}{\vkap p_{2}}\gamma^{\mu} + \gamma^{\mu}\frac{\slashed{\vkap}\slashed{a}}{\vkap p_{1}} \Big) \nonumber \\
&& \qquad\qquad+ C_{3} \frac{e^{2}\slashed{a}\slashed{\vkap}\gamma^{\mu}\slashed{\vkap}\slashed{a}}{4\vkap p_{2}\vkap p_{1}} \Big]\frac{u_{\sigma_{1}}(p_{1})}{\sqrt{2p_{1}^{0}V}}, \label{eqn:GamStruc}
\eea
\bea
C_{n}(r,c_{2},c_{3}) &\defto& \int_{-\infty}^{\infty} d\vphi~ \vphi^{n-1} \mbox{e}^{i ( r\vphi + c_{2} \vphi^{2} + c_{3} \vphi^{3})}. \label{eqn:Cn}
\eea
By shifting the $\vphi$ co-ordinate $\vphi\to \vphi-c_{2}/3c_{3}$, one can show:
\bea
C_{1} &=& b\,\Ai(\mu^{2/3})\label{eqn:I0a}\\
C_{2} &=& -b\left[\frac{c_{2}}{3c_{3}}\Ai(\mu^{2/3})+\frac{i}{(3c_{3})^{1/3}}\Ai'(\mu^{2/3})\right]\\
C_{3} &=& b\,\Bigg\{\!\left[\Bigg(\frac{c_{2}}{3c_{3}}\Bigg)^{2}-\left(\frac{\mu}{3c_{3}}\right)^{2/3}\right]\Ai(\mu^{2/3})\nonumber \\
&&  \qquad\qquad +\frac{2ic_{2}}{(3c_{3})^{4/3}} \Ai'(\mu^{2/3})\Bigg\},\label{eqn:I0b}
\eea
\begin{gather}
b= \frac{2\pi}{(3c_{3})^{1/3}}\mbox{e}^{i\eta};\qquad \eta = -\frac{rc_{2}}{3c_{3}} +\frac{2c_{2}^{3}}{27c_{3}^{2}}; \nonumber\\ \mu^{2/3}= \frac{r-c_{2}^{2}/3c_{3}}{(3c_{3})^{1/3}}, \label{eqn:Cdefs}
\end{gather}
where $\Ai$ is the Airy-function defined in \eqnref{eqn:Airy} with normalisation $N=\pi$.
\newline

The calculation of $\Delta_{\mu}(s)$ proceeds in a similar way. From \eqnref{eqn:Del1} we have:
\bea
\Delta_{\mu}(s) &=& \int\! d\vphi \left\{\frac{\overline{u}_{\sigma_{3}}(p_{3})}{\sqrt{2p_{3}^{0}V}} \Big[1+\frac{e\slashed{A}(\vphi)\slashed{\vkap}}{2\vkap p_{3}} \Big]\gamma_{\mu}\Big[1-\frac{e\slashed{\vkap}\slashed{A}(\vphi)}{2\vkap p_{4}}\Big]\right. \nonumber 
\\ && \left. \frac{v_{\sigma_{4}}(p_{4})}{\sqrt{2p_{4}^{0}V}}\mbox{e}^{i ( s\vphi + c_{2}' \vphi^{2} + c_{3}' \vphi^{3})}\right\},
\eea
where we have introduced:
\begin{gather}
c_{2}' \defto \frac{e}{2}\Big(\frac{p_{3}a}{\vkap p_{3}} - \frac{p_{4}a}{\vkap p_{4}}\Big) , ~~ c_{3}' \defto -\frac{e^{2}a^{2}}{6}\Big(\frac{1}{\vkap p_{3}} + \frac{1}{\vkap p_{4}}\Big). \label{eqn:alphabetap}
\end{gather}
Then:
\bea
\Delta_{\mu}(s) &=& \frac{\overline{u}_{\sigma_{3}}(p_{3})}{\sqrt{2p_{3}^{0}V}} \Big[D_{1}\gamma_{\mu}+D_{2}\frac{e}{2}\Big(\frac{\slashed{a}\slashed{\vkap}}{\vkap p_{3}}\gamma_{\mu} - \gamma_{\mu}\frac{\slashed{\vkap}\slashed{a}}{\vkap p_{4}} \Big) \nonumber \\ && - D_{3}\frac{e^{2}\slashed{a}\slashed{\vkap}\gamma_{\mu}\slashed{\vkap}\slashed{a}}{4\vkap p_{3}\vkap p_{4}} \Big]\frac{v_{\sigma_{4}}(p_{4})}{\sqrt{2p_{4}^{0}V}}, \label{eqn:Ddefs}
\eea
\bea
D_{n} &=& C_{n}(r \to s, c_{2}\to c_{2}', c_{3}\to c_{3}').
\eea

\subsection{Fermion trace} \label{app:trace}

The trace to be evaluated comprises the trace of each exchange term mod-squared plus interference terms. If we define the objects: 
\begin{gather}
\widetilde{M}_{\mu} =\gamma^{0} M_{\mu}^{\dagger} \gamma^{0}; \qquad
\mathcal{E}^{\pm}_{i} = \frac{\pm\slashed{p}_{i} + m}{2m},
\end{gather}
then for each exchange term squared, the trace is of the form:
\begin{align}
\begin{split}
\mathcal{I}(\rightarrow,\rightarrow) &= \sum_{\sigma_{i}}\!\!\tr\!\!\Big[\overline{u}_{\sigma_{2}} C^{\mu}(p_{2},p_{1},r) u_{\sigma_{1}}\overline{u}_{\sigma_{3}} D_{\mu}(p_{3},p_{4},r) v_{\sigma_{4}}\\
&   u_{\sigma_{2}} \widetilde{C}^{\nu}(p_{2},p_{1},r') \overline{u}_{\sigma_{1}} u_{\sigma_{3}} \widetilde{D}_{\nu}(p_{3},p_{4},r') \overline{v}_{\sigma_{4}} \Big]\\
&= -\tr \Big[\mathcal{E}^{+}_{1} C^{\mu}(p_{2},p_{1},r) \mathcal{E}^{+}_{2} C^{\dagger\,\nu}(p_{2},p_{1},r')\Big]\\
& .\,\tr\Big[ \mathcal{E}^{+}_{3} D_{\mu}(p_{3},p_{4},r) \mathcal{E}^{-}_{4}  D^{\dagger}_{\nu}(p_{3},p_{4},r')\Big],
\label{eqn:Irr}
\end{split}
\end{align}
where from the definition of $\mathcal{I}(\rightarrow,\rightarrow)$ \eqnref{eqn:Sfi4d}, $C^{\mu}$ and $D^{\mu}$ are factors of $C_{j}$ and $D_{j}$, $j\in\{1,2,3\}$, multiplied by the combinations of gamma matrices occurring in \eqnrefs{eqn:Cdefs}{eqn:Ddefs}, integrated over $r$ and $r'$ variables. Following similar steps, for each interference term it is of the form:
\begin{align}
\begin{split}
\mathcal{I}(\leftarrow,\rightarrow) &= \sum_{\sigma_{i}}~\tr\,\Big[\overline{u}_{\sigma_{3}} C^{\mu}(p_{3},p_{1},r) u_{\sigma_{1}}\overline{u}_{\sigma_{2}} D_{\mu}(p_{2},p_{4},r) v_{\sigma_{4}}\\
& u_{\sigma_{2}}\widetilde{C}^{\nu}(p_{2},p_{1},r') \overline{u}_{\sigma_{1}} u_{\sigma_{3}} \widetilde{D}_{\nu}(p_{3},p_{4},r') \overline{v}_{\sigma_{4}}\Big]\\
&=  -\tr\,\Big[\mathcal{E}^{+}_{1} C^{\dagger\, \nu}(p_{2},p_{1},r') \mathcal{E}^{+}_{2}D_{\mu}(p_{2},p_{4},r) \\
& \mathcal{E}^{-}_{4}D^{\dagger}_{\nu}(p_{3},p_{4},r')\mathcal{E}^{+}_{3} C^{\mu}(p_{3},p_{1},r)\Big]. 
\label{eqn:Ill}
\end{split}
\end{align}
We note that the trace of each exchange term mod-squared is factorisable into a Compton-scattering and pair-creation vertex when the connecting photon polarisation is taken into account. These traces were performed with the package {\tt Feyncalc} \cite{feyncalc}.

\subsection{Complex phase factor}
It can be seen from the definitions of the Airy integrals resulting from the vertex factors, that an overall phase factor $\eta(\cdot,\cdot)$ occurs in the traces $\mathcal{I}(\cdot,\cdot)$ (from the $\eta$ factors in \eqnref{eqn:Cdefs} occurring in \eqnrefs{eqn:Irr}{eqn:Ill}). If one squares the $r$ integral, labelling the new co-ordinate $r'$, they are of the form:
\begin{align}
\begin{split}
\eta(\ra,\ra) &= -\left(\frac{c_{2}(p_{2},p_{1})}{3c_{3}(p_{2},p_{1})}-\frac{c_{2}'(p_{3},p_{4})}{3c_{3}'(p_{3},p_{4})}\right)(r-r')\label{eqn:eta0}\\
\eta(\ra,\la) &= -\frac{(r+r_{\ast})c_{2}(p_{2},p_{1})}{3c_{3}(p_{2},p_{1})}+\frac{(r-s_{\ast})c_{2}'(p_{3},p_{4})}{3c_{3}'(p_{3},p_{4})}  +\\
& \frac{(r'+r_{\ast})c_{2}(p_{3},p_{1})}{3c_{3}(p_{3},p_{1})}-\frac{(r'-s_{\ast})c_{2}'(p_{2},p_{4})}{3c_{3}'(p_{2},p_{4})}  + \\
& \frac{2}{27}\left(\frac{c_{2}^{3}(p_{2},p_{1})}{c_{3}^{2}(p_{2},p_{1})} -\frac{c_{2}^{3}(p_{3},p_{1})}{c_{3}^{2}(p_{3},p_{1})}+\frac{c_{2}^{\prime\, 3}(p_{3},p_{4})}{c_{3}^{\prime\,2}(p_{3},p_{4})}\right. \\
&\left. -\frac{c_{2}^{\prime\,3}(p_{2},p_{4})}{c_{3}^{\prime\,2}(p_{2},p_{4})}\right).
\end{split}
\end{align}
After some simplification, it can be seen that the non-interference terms have a relatively simple structure (where $\chi_{E}=E/\Ecr$, $\Ecr=m^{2}/e$):
\bea
\eta(\ra,\ra) &=& \frac{-\vkap^{0}(r-r')}{m^{2}\chi_{E}(p_{1}^{-}-p_{2}^{-})} \left[p_{2}^{1}(p_{3}^{-}-p_{1}^{-}) + p_{3}^{1}(p_{1}^{-} - p_{2}^{-}) \right.\nonumber\\
&&  \left. \qquad + p_{1}^{1}(p_{2}^{-}-p_{3}^{-})\right].
\eea

\subsection{Formation length of two-step and sub-processes} \label{app:temp}
If we imagine that the process takes place in a coherence interval of finite duration, for a given incoming electron momentum $p_{1}$, these integrals will also be finite (otherwise the particles would have to be accelerated infinitely quickly (see also \cite{ilderton12a})). We apply the following reasoning, which is standard for lower-order constant-crossed-field processes (see e.g. \cite{ritus85}). The phase of the modified Airy functions that occur at each vertex \eqnref{eqn:Cn} have a stationary point at:
\bea
\varphi^{\ast} = -\rho\left[1\pm \sqrt{1-\frac{3rc_{3}}{c_{2}^{2}}}\right]; \qquad \rho = \frac{c_{2}}{3c_{3}}.
\eea 
Let us write $\varphi^{\ast}=-\rho(1\pm\delta\varphi^{\ast})$, where $\rho$ is the phase at which the process takes place (the saddle-point) and $\delta\varphi^{\ast}$ is the width. It is assumed that $\delta \varphi^{\ast}=0$ because when this is not the case, the resulting Airy functions have no dependency on dynamical variables. Therefore, integration over $\rho$ is equivalent to integration over the relevant part of the phase.

The complex phase factor for the purely direct (and analogously for the purely exchange) term in a single $r$-integral is of the form:
\bea
\eta(\ra,\ra)=(\varphi^{\ast}_{x}-\varphi^{\ast}_{y})r,
\eea
where  $\varphi^{\ast}_{z}$ is the saddle-point at co-ordinate $z$ and for the purely direct term
\bea
\varphi^{\ast}_{x} &=& \frac{\vkap^{0}}{m^{2}\chi_{E}}\frac{p_{2}^{1}p_{1}^{-} -p_{1}^{1}p_{2}^{-}}{p_{1}^{-}-p_{2}^{-}} \label{eqn:phix}\\
\varphi^{\ast}_{y} &=& \frac{\vkap^{0}}{m^{2}\chi_{E}}\frac{p_{2}^{1}p_{3}^{-} +p_{3}^{1}(p_{1}^{-}-p_{2}^{-})-p_{1}^{1}p_{3}^{-}}{p_{1}^{-}-p_{2}^{-}}.\label{eqn:phiy}
\eea

\subsection{Isolation of the two-step process} \label{app:12removal}
Concentrating on the non-exchange term $\mathcal{I}(\rightarrow,\rightarrow)$ (an analogous calculation follows for $\mathcal{I}(\leftarrow,\leftarrow)$), the integral over $r$ is of the form: 
\begin{align}
\mathcal{J}' = \frac{1}{\pi^{2}}\int\!\!dp_{2}^{1}\, dp_{3}^{1}\,\left|\int\!dr~\frac{\mbox{e}^{i[\varphi^{\ast}_{x}(p_{2}^{1})-\varphi^{\ast}_{y}(p_{2}^{1},p_{3}^{1})]r}F(r)}{(r+i\eps)}\right|^{2}\label{eqn:sep0}
\end{align}
with $F(r)\in\mathbb{C}$. Performing an integral substitution $\varphi_{\pm}^{\ast} = \varphi_{x}^{\ast}\pm\varphi_{y}^{\ast}$, one can rewrite this as:
\bea
\mathcal{J}' = \frac{1}{2J\pi^{2}}\int d\varphi^{\ast}_{+}\, d\varphi^{\ast}_{-}\,\left|\int\!dr~\frac{\mbox{e}^{i\varphi^{\ast}_{-}r}F(r)}{(r+i\eps)}\right|^{2},
\eea
where $J=|\partial(\varphi^{\ast}_{+},\varphi^{\ast}_{-})/\partial (p_{2}^{1},p_{3}^{1})|$ is the inverse Jacobian:
\bea
J = m^{2}\xi^{2} \,\frac{p_{1}^{-}-p_{2}^{-}}{2p_{1}^{-}}.
\eea
In order to remain consistent, before integrating in the variable $\varphi^{\ast}_{-}$, we will first perform the principal value calculation. The order of integration is important as principle value and $\varphi^{\ast}_{+,-}$ integrals do not necessarily commute (for example in \eqnref{eqn:PV1}, integration in $a$ does not commute with the operation $\hat{\mathcal{P}}$).
\begin{widetext}
\begin{align}
\mathcal{J}'&= \frac{1}{2J\pi^{2}}\int d\varphi^{\ast}_{+}\, d\varphi^{\ast}_{-} \left|-i\pi F(0) + \hat{\mathcal{P}}\int\!dr~\frac{\mbox{e}^{i\varphi^{\ast}_{-}r}F(r)}{r}\right|^{2}\\
&= \frac{1}{2J\pi^{2}}\int d\varphi^{\ast}_{+}\, d\varphi^{\ast}_{-}\,\left|-2i\pi F(0)\theta(-\varphi^{\ast}_{-}) + \int\!dr~\mbox{e}^{i\varphi^{\ast}_{-}r}~\frac{F(r)-F(0)}{r}\right|^{2}\label{eqn:sep2}\\
&= \frac{1}{J}\left\{2|F(0)|^{2}\int d\varphi^{\ast}_{+}\, d\varphi^{\ast}_{-}\,\theta(-\varphi^{\ast}_{-})+ \frac{1}{\pi}\int\!d\varphi^{\ast}_{+}\int_{-\infty}^{\infty}\!dr~\frac{|F(r)-F(0)|^{2}}{r^{2}}\right.\nonumber\\
& \left.+\frac{1}{\pi}\left[ F(0) \int\!d\varphi^{\ast}_{+}\int_{0}^{\infty} \!dr~\frac{F^{\ast}(r)+F^{\ast}(-r)-2F^{\ast}(0)}{r^{2}} + \trm{ c. c.}\right]\right\} \label{eqn:sep1},
\end{align}
\end{widetext}
where c. c. stands for complex conjugate and in \eqnref{eqn:sep2} and \eqnref{eqn:sep1} respectively, we have used the results \cite{heitler60}:
\begin{gather}
\hat{\mathcal{P}}\int_{-\infty}^{\infty}\!\frac{dr}{r}\,\mbox{e}^{iar} = i\pi\,\trm{sgn}(a) \label{eqn:PV1}\\
\int_{-\infty}^{\infty}\!d\varphi\,\theta(\varphi)\,\mbox{e}^{i\varphi r} = i\hat{\mathcal{P}}\,\frac{1}{r} + \pi \delta(r).
\end{gather}
The first term in \eqnref{eqn:sep1} can be identified as the two-step process due to the two phase integrals that occur, justified in the main text. In addition, a Heaviside theta-function in $-\varphi^{\ast}_{-}$ occurs, which is a sign that causality is preserved insofar as pair-creation can only occur after Compton scattering in this setting. 

\subsection{Justification for neglecting interference terms between direct and exchange parts}\label{app:crossremoval}
For the non-interference terms, it has been shown in \eqnreft{eqn:sep0}{eqn:sep1} how the simple nature of the exponential occurring in the vertex functions leads to a dependency on the external field phase in $\Delta\varphi$. For the interference terms the integral over $p_{2,3}^{1}$ is of the form
\bea
\mathcal{J}'(\ra,\la)\!=\!\int\!\! dp_{2}^{1}\, dp_{3}^{1}\,dr\,dr'\frac{\mbox{e}^{i\eta(\rightarrow, \leftarrow)}F(r)F^{\ast}(r')}{(r+i\eps)(r'-i\eps)},\nonumber \\ \label{eqn:cross1}
\eea
where the phase $\eta(\ra, \la)$ contains terms of the order $p_{2,3}^{1}$, $(p_{2,3}^{1})^{3}$, $p_{2}^{1}(p_{3}^{1})^{2}$ and $(p_{2}^{1})^{2}p_{3}^{1}$. Instead of generating an arbitrarily large (divergent in the strict sense) factor of formation length $\Delta\varphi$ as for each exchange term squared, for this interference term, an Airy function in remaining particle momenta and $r$ is generated, which, having positive or negative argument, will tend to reduce the value of the integral. As noted in the conclusions of the main text, if one demands that $\xi\Delta\varphi\gg1$ to prevent regions of negative total probability from arising, then these interference terms can be safely neglected.


\subsection{Numerical evaluation of one-step integral} \label{app:osi}
Despite the five-dimensional integration of $\mathcal{I}^{(1)}$ \eqnref{eqn:I1int} not being oscillatory, it is challenging to numerically evaluate
\bea
\mathcal{I}^{(1)} &=& \frac{\alpha^{2}}{\pi\chi_{1}}\!\!\int\!\! \frac{d\chi_{2}d\chi_{3}dp_{2}^{2}dp_{3}^{2}dv\theta(\chi_{1}-\chi_{2}-\chi_{3})}{(\chi_{1}-\chi_{2})^{2}} \frac{\mathcal{B}^{(1)}}{v^{2}}. \label{eqn:I1int}
\eea
As the evaluation of $\mathcal{B}^{(1)}$ is computationally expensive, it is important to know the relevant bounds of the variables. In the main text, it was justified that $p_{2,3}^{2}\approx 1$. We then take $p_{2}^{2}, p_{3}^{2}, \in [-4, 4]$. From the arguments of the Airy function given in \eqnref{eqn:nus}, one would expect the maximum of $\mathcal{B}^{(1)}$ in $v$ to be of the order $v\approx \chi_{1}\mu_{2,3}$. Let $a=\chi_{2}/\chi_{1}$ and $b=\chi_{3}/\chi_{1}$ so that $a$, $b$ $\in [0,1]$, then
\bea
\chi_{1}\mu_{2} = \frac{1}{a}-1;\quad \chi_{1}\mu_{3} = \frac{1}{1-a-b}+\frac{1}{b}.
\eea
However, from studies of the approximated two-step process in \cite{king13a}, it seems that as $\chi_{1}$ increases above $1$, $(1-a)/a \sim 1/\chi_{1}$. Likewise, $\chi_{3}$ was observed to remain approximately constant so that $b \to 1/\chi_{1}$, leading to $\chi_{1}\mu_{2,3}\sim\chi_{1}$. Therefore, $v \in [0,10\chi_{1}]$ was chosen for the $v$ integration, with the tail $[10\chi_{1},\infty]$ evaluated in $w$ with the conformal transformation $w=\tan^{-1}v$. Although the function $\mathcal{B}$ is quite smooth in the $\chi_{2}$--$\chi_{3}$ plane, the largest contribution to the total integral originates from an ever-smaller region around $a=1$, $b=0$, with increasing $\chi_{1}$, making these points particularly costly to evaluate. To escape the triangular $\chi_{2}$--$\chi_{3}$ plane as given in Fig. 2 in the main text, one can substitute integration variables $\chi_{2}\to\chi_{2}/(\chi_{1}-\chi_{3})$ and $\chi_{3}\to\chi_{3}/\chi_{1}$ to achieve a square integration region between $0$ and $1$. One can 
then easier observe where the maxima lie in the integrand and evaluate grids of points incorporating these. The resulting surface can then be interpolated and numerically integrated. 
\begin{figure}[!h]
\centering\noindent
  \includegraphics[draft=false, width=7.4cm]{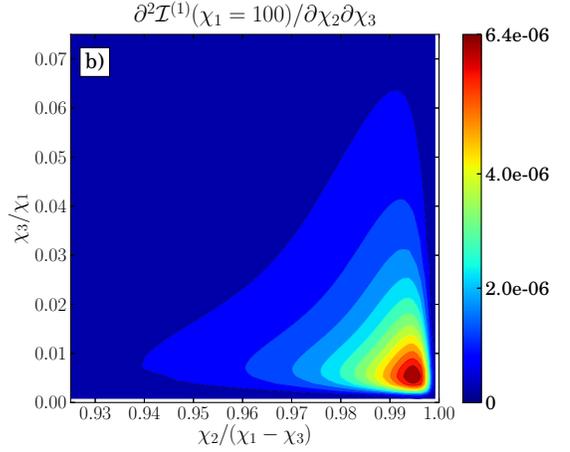}\hfill
\caption{The differential of the dynamical part of the one-step rate $\partial\mathcal{I}^{(1)}/\partial \chi_{2}\partial\chi_{3}$ for $\chi_{1}=100$.} \label{fig:chi2chi3}
\end{figure}
For $\chi>1$, The global maximum was found in an ever smaller region $\chi_{3}/\chi_{1}\to 0$, $\chi_{2}/(\chi_{1}-\chi_{3})\to1$. For example, for the case $\chi_{1}=100$, as plotted in \figref{fig:chi2chi3}, the maximum was centred around $\chi_{2}/(\chi_{1}-\chi_{3})=0.995$, $\chi_{3}/\chi_{1}=0.005$.
The tests of accurate integration, other than variation of number of points and integration region, were provided by comparison with asymptotic expressions from the literature for the total rate and the Weizs\"acker-Williams approximation.

\section{Integrals of Airy functions} \label{app:Airy}
We give here a selection of Airy integrals that are useful in the derivation and are in part derived from other results in the literature. Let us define:
\bea
I_{2n} &=& \int_{-\infty}^{\infty}dt~ t^{2n}\Ai^{2}(t^{2} + c)\\
J_{2n} &=& \int_{-\infty}^{\infty}dt~ t^{2n}\Ai(t^{2} + c)\Ai'(t^{2} + c)\\
K_{2n} &=& \int_{-\infty}^{\infty}dt~ t^{2n}\Ai^{\prime\,2}(t^{2} + c),
\eea
where $c$ is an arbitrary constant (integrals involving coefficients with odd powers of $t$ are zero due to the functions being odd). With $I_{0}$, $J_{0}$ and $K_{0}$ being given in e.g. \cite{aspnes66}, from partial integration and the use of some primitives given in \cite{albright77}, the following analytical results have also been verified numerically:
\bea
I_{0} &=& \frac{\pi}{2N} ~\Ai_{1}(v)\\
I_{2} &=& -\frac{\pi}{4N} ~\left[\frac{1}{\kappa}\Ai'(v)+c\,\Ai_{1}(v)\right]\\
I_{4} &=& \frac{3\pi}{16N}~\left[\frac{\kappa}{4}\Ai(v) + \frac{c}{\kappa}\Ai'(v) + c^{2}\Ai_{1}(v) \right]\\
J_{0} &=& -\frac{\kappa \pi}{4N }\Ai(v)\\
J_{2} &=& -\frac{\pi}{8N}\Ai_{1}(v)\\
J_{4} &=& \frac{3\pi}{16N}\left[\frac{1}{\kappa}\Ai'(v) + c \Ai_{1}(v)\right]\\
K_{0} &=& -\frac{\pi}{4N \kappa}\left[3\Ai'(v) + v \Ai_{1}(v) \right]\\
K_{2} &=& \frac{\pi}{16N}\left[\frac{5}{4}\kappa \Ai(v) + \frac{c}{\kappa}\Ai'(v) + c^{2} \Ai_{1}(v)\right],
\eea
$v=\kappa c$, $\kappa = 2^{2/3}$, $N$ is the normalisation factor occuring in the definition of the Airy function:
\bea
\Ai(x) \defto \frac{1}{N} \int_{0}^{\infty} dt~ \cos \left(t^{3} + xt\right) \label{eqn:Airy}
\eea
where $\Ai'(x) = \partial\Ai(x)/\partial x$ and $\Ai_{1}$ is defined as:
\bea
\Ai_{1}(x) = \int_{0}^{\infty} dt ~\Ai(t+x). \label{eqn:Airy1}
\eea

\bibliography{current}

\end{document}